\documentclass[%
reprint,
nofootinbib,
amsmath,amssymb,
aps,
pre,
]{revtex4-2}

\usepackage{xcolor}
\usepackage{graphicx}
\usepackage{dcolumn}
\usepackage{bm}
\usepackage{soul}
\usepackage{natbib}
\usepackage{comment}

\usepackage[mathlines]{lineno}

\newcommand{\dd}{\mathrm{d}}
\newcommand{\ff}{\mathrm{f}}
\newcommand{\jj}{\mathrm{j}}
\newcommand{\bb}{\mathrm{b}}
\newcommand{\cc}{\mathrm{c}}
\newcommand{\gx}{\mathrm{g}}
\newcommand{\pp}{\mathrm{p}}

\begin{document}

\allowdisplaybreaks

\title{Power laws and phase transitions in heterogenous car following with reaction times  }
\author{A. Sai Venkata Ramana}
\email{\url{Email: sa183@nyu.edu} }
\affiliation{New York University Abu Dhabi, Saadiyat Island, P.O. Box 129188, Abu Dhabi, United Arab Emirates}

\author{Saif Eddin Jabari}
\email{Corresponding author \url{ Email: sej7@nyu.edu} }
\affiliation{New York University Abu Dhabi, Saadiyat Island, P.O. Box 129188, Abu Dhabi, United Arab Emirates}
\affiliation{New York University Tandon School of Engineering, Brooklyn New York 11201, United States of America}
\begin{abstract}
We study the effect of reaction times on the kinetics of relaxation to stationary states and on congestion transitions in heterogeneous traffic using simulations of Newell’s model on a ring. Heterogeneity is modeled as quenched disorders in the parameters of Newell’s model and in the reaction time of the drivers.We observed that at low densities, the relaxation to stationary state from a homogeneous initial state is governed by the same power laws as derived by E. Ben-Naim et al., Kinetics of clustering in traffic flow, Phys. Rev. E 50, 822 (1994). The stationary state, at low densities, is a single giant platoon of vehicles with the slowest vehicle being the leader of the platoon. We observed formation of spontaneous jams inside the giant platoon which move upstream as stop-go waves and dissipate at its tail. The transition happens when the head of the giant platoon starts interacting with its tail, stable stop-go waves form, which circulate in the ring without dissipating. We observed that the system behaves differently when the transition point is approached from above than it does when approached from below. When the transition density is approached from below, the gap distribution behind the leader has a double peak and is fat-tailed but has a bounded support and thus the maximum gap in the system and the variance of the gap distribution tend to size-independent values. When the transition density is approached from above, the gap distribution becomes a power law and, consequently, the maximum gap in the system and the variance in the gaps diverge as a power law, thereby creating a discontinuity at the transition. Thus, we observe a phase transition of unusual kind in which both a discontinuity and a power law are observed at the transition density. These unusual features vanish in the absence of reaction time, i.e., when the vehicles react instantaneously to a perturbation ahead (e.g., automated driving). Overall, we conclude that the nonzero reaction times of drivers in heterogeneous traffic significantly change the behavior of the free flow to congestion transition while it doesn’t alter the kinetics of relaxation to stationary state.

\end{abstract}

\keywords{Traffic flow, power laws, phase transitions, reaction time, quenched disorder}
\maketitle

\section{Introduction}
Traffic systems are nonequilibrium driven systems that exhibit rich collective phenomena in the kinetics of relaxation to a nonequilibrium stationary state and in the phase transition to the congestion regime.  The problem becomes even more complex when a quenched disorder is introduced into the system. A quenched disorder, in a physical sense, implies heterogeneous traffic in which each driver-vehicle unit has a different set of parameters.  At a fundamental level, two basic kinds of approaches are used to model traffic: car-following models \cite{Helbing2001} and cellular automata \cite{DC2000}. Physicists have studied the effects of quenched disorder on the collective phenomena using the totally asymmetric simple exclusion process (TASEP) and the Nagel-Schkrekenberg (NS) models \cite{stocNS1992,Nagel1993,Chowdhury1997}, which are cellular automata, while such studies are rare in the car-following literature. 
At the outset it might appear that the conclusions of the car-following model would be the same as those in the cellular automata. However, it is important to understand if the subtle differences in basic assumptions   between the methods might lead to differences in the emergent phenomena. It is also important to characterize these phenomena in the jargon of car-following methods, which are widely used in transportation engineering applications~\cite{Nagel,Sumo,Vissim}. In previous work, we studied heterogeneous traffic by introducing quenched disorders into the parameters of Newell's model \cite{Ramana2020}. In this work, we study the further effect of the reaction time of drivers by means of numerical simulations of Newells'  model.  
We observe that the presence of reaction times and quenched disorders in them significantly changes the picture from one with no reaction times and reveals interesting phenomena that are not observed in the cellular automata. 
We first give a brief review of related works that use cellular automata, specifically the NS model in Sec. II and highlight points that prompted the present study. 
In Sec. III, we present our car following model and explain briefly the role played by reaction times in the formation of spontaneous stop-go waves.  In Sec. IV, we discuss the approach to stationary state. In Sec. V, the giant platoon that forms in the  stationary state is characterized. In Sec. VI, the phase transition from the platoon forming phase to the congestion phase is analyzed. The results are summarized in Sec. VII.

\section{Collective Phenomena in NS model}
The NS model for a one dimensional lattice of $L$ sites on a ring is as follows: The speed of each vehicle is assumed to be discrete with allowed integer values between $0$ and $v_{\max}$. The time step is taken to be unity and dimensionless. Thus, the space gap ($d$) and speed ($v$) have the same units. Starting from a given initial configuration, the positions and speeds of all the vehicles are updated at each time step according to the following rules \cite{DC2000}: (i) The speed $v_i$ of the $i$th vehicle is updated to $\min\{v_{\max}, v_i +1\}$ if $v_i < d_i$, where $d_i$ is the gap ahead of the $i$th vehicle. (ii) If $v_i \ge d_i$, $v_i$ is updated to $d_i-1$. (iii) The speed of a vehicle is reduced by unity ($v_i \mapsto v_i-1$) with a probability $p$ to account for randomness in hopping, also called as random deceleration. (iv) Each vehicle advances $v_i$ sites. The NS model is very similar to TASEP except that in TASEP $v_{\max}=1$ and the positions of particles are updated in a random sequential manner.

The traffic system as described by the TASEP or the NS model is intrinsically a driven non-equilibrium system.
The evolution of such system towards its stationary state reveals its dynamical universality class which may be distinguished based on the dynamical exponent $z$ related to the emerging length scale $\xi(t)$ in the system as $\xi \sim t^{1/z}$. In the case of traffic, $\xi(t)$ is the length of the platoon of vehicles moving as a cluster. While there have been early numerical studies in the NS literature regarding the dynamical universality class of the NS  model \cite{Csanyi1995,Sasvari1997}, it has been only recently proven by Gier et al.~\cite{Gier2019} using non-linear fluctuating hydrodynamics that the NS model belongs to the super-diffusive Kardar-Parisi-Zhang universality class with dynamical exponent $z=3/2$. 

There has been intense debate in the literature over the nature of the jamming phase-transition and the  corresponding order parameter for the NS model  \cite{Lubeck1998,Roters1999,DC2000comment,Roters2000reply,Eisenblatter1998}.  Some studies used the observation of a double peak in the distribution functions of gap and speed to identify the critical density~\cite{Chowdhury1997,Lubeck1998,Roters1999, Bain2016,Souza2009}.  Gerwinsky et al.~\cite{Gerwinski1999}
noted that the jams can be observed at any density and argued the formation of a stable jam as a criterion for the phase transition.  Recently, Bette et al.~\cite{Bette2017} decomposed the jams based on the mechanism of their formation and observed that the formation of stable jams lead to the phase transition while the unstable jams can be present at any density. It may be noted that the basic reason for the formation of jams is the stochasticity induced by $p$.

The presence of a quenched disorder in the system makes it even more complex. Krug and Ferrari~\cite{Krug1996,Krug2000} studied a version of TASEP with quenched disorder in $p$ with probability distribution $f(p) \sim (p-p^{\min})^n$  when $p \rightarrow p^{\min}$ and conjectured that the dynamical exponent $z$ depends on the exponent of the quenched disorder as $z=(n+2)/(n+1)$. Krug and Ferrari also argued that the phase transition from platoon forming phase to a laminar phase without platoon formation would be of second order if $n \leq 1$ and first order for $n>1$. Evans~\cite{Evans1996} independently solved for the steady state of the TASEP with quenched disorder in jump rates and showed  that the phenomenon of bunching of vehicles behind the slowest vehicle is analogous to the Bose-Einstein condensation.  Ktitarev et al.~\cite{Ktitarev1997}  did simulations of the NS model with quenched disorder ($f(p)$) in $p$ and concluded that the dynamical exponent $z$ and the exponent for the gap distribution near the critical point are same as those conjectured by Krug and Ferrari. Bengrine et al. ~\cite{Bengrine1999} simulated NS model for open boundary conditions. Their conclusions also corroborated Krug's conjectures regarding the order of the transition and the exponent $z$. 

It can be seen from the above discussion that the stochasticity induced by the random deceleration (which occurs with probability $p$ as discussed above) plays a central role in the collective phenomena exhibited by the NS model. The random hopping probability $p$ has been introduced to account for spontaneous traffic jams (also called stop-go waves) and various other aspects like non-deterministic acceleration by drivers etc. However, it is not directly related to
 any physically observable phenomenon in traffic flow. There is no analogous parameter in car-following models also as can be seen in Newell's model explained in the next section.  The main effect of the random deceleration in the NS model (i.e., the formation of stop-go waves) is captured by driver reaction times in car-following models, and this basically models the delay in response of a driver-vehicle unit to a perturbation ahead of it. The process by which the stop-go waves form due to reaction times has its origin in the flow instability, which is a deterministic process and thus could be quite different from the way stop-go waves form in the NS model, which is based on stochasticity induced by $p$. This raises the question of whether the collective phenomena exhibited by car-following models would be the same as those observed in the NS model. Our present study aims to address this point.

\section{Car-following model for heterogeneous traffic }
\begin{figure}
	\resizebox{0.45\textwidth}{!}{%
\includegraphics[angle=0]{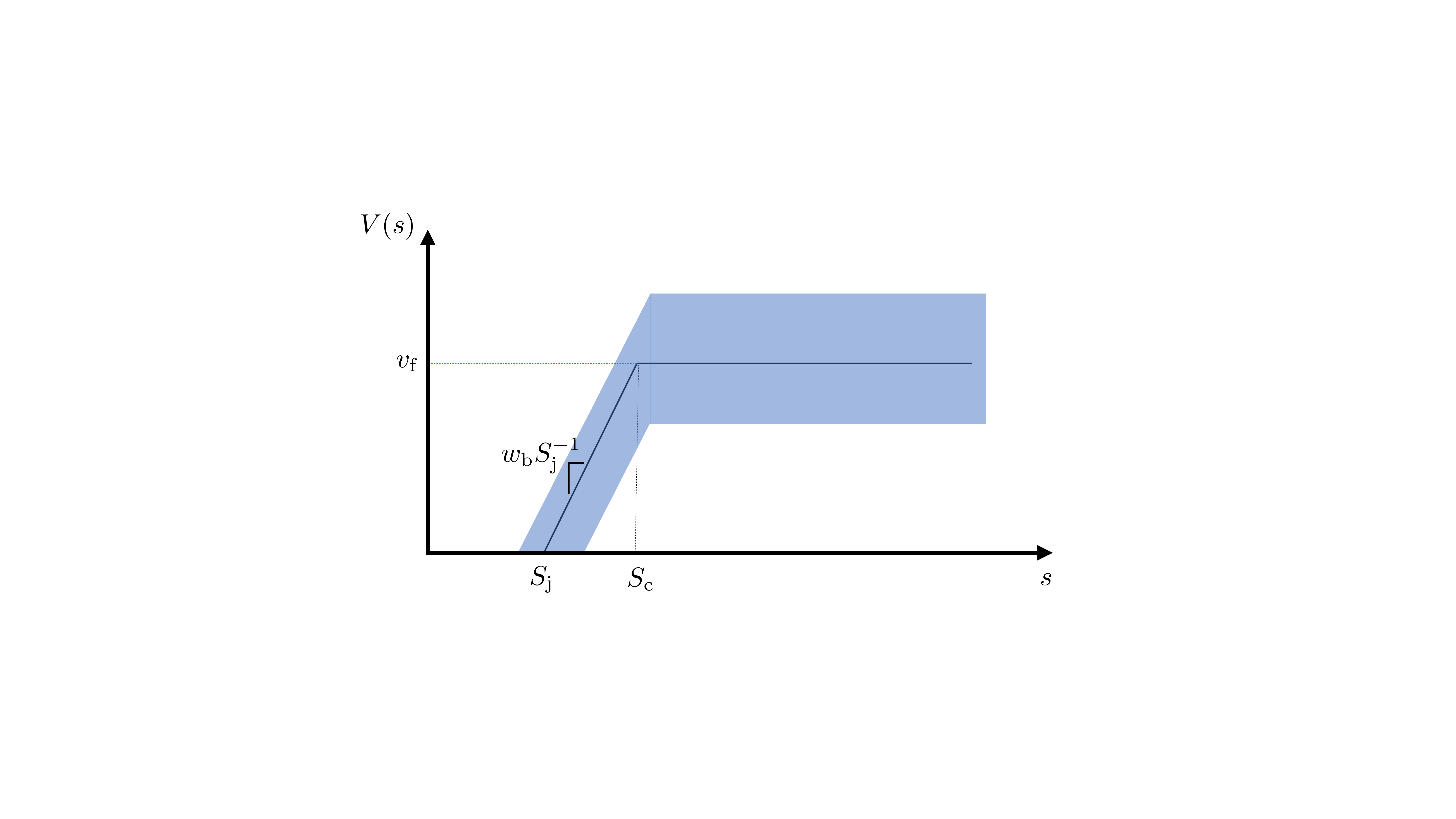}}
\caption{A schematic of the speed $V(s)$ versus gap $s$ profile of a driver as per Newell's model. A typical profile is shown by black line. Below jam gap $S_{\jj}$, the vehicle is at rest. If the gap $s$ is between $S_{\jj}$ and $S_{\cc}$,
the speed increases at a rate of $w_{\bb} S_{\jj}^{-1}$ and if the gap $s$ is above $S_{\cc}$,  the speed is maximum and constant at $v_{\ff}$. The filled blue area encompasses various possible driver profiles in the system as all the parameters of the model are random in our present study.} \label{fig:schN}

\end{figure}

Newell's model~\cite{Newell2002} is a simple physical car-following model that is known to reasonably capture the dynamics of car-following.  It has been empirically validated in a number of studies \cite{ahn2004verification,chiabaut2009fundamental,chiabaut2010heterogeneous,jabari2014}. The equation of motion for an $i$th vehicle in the model is 
\begin{equation}
    \frac{\dd x_i(t)}{\dd t} = V(s_i(t-\tau_i)), \label{EOM}
\end{equation}
where $x_i(t)$ is the position of vehicle $i$ at time $t$ and $V(\cdot)$ is a speed relation that takes the spacing between vehicle $i$ and their leader ($s_i \equiv x_{i-1} - x_i$) as input. The assumptions of Newell's model are embedded in $V$, which basically couples the dynamics of the $i^{th}$ vehicle with that of its leader, the $(i-1)^{th}$ vehicle, and accounts for $i$'s reaction time, denoted by $\tau_i$.  The reaction time may be interpreted as the delay in response of the driver to a perturbation ahead. A schematic of the speed versus gap relation is shown in Fig.~\ref{fig:schN} Assuming no overtaking, $x_{i-1}>x_i$ at all times. The equation above maintains total asymmetry in the interactions as the $i$th vehicle only interacts with the vehicle ahead but not vice versa. Newell's conjecture for speed is as follows:
\begin{equation}
	V(s) = \begin{cases} v_{\ff}, & s \ge S_{\cc} \\ w_{\bb} \max \{S_{\jj}^{-1}s-1,0\}, &s < S_{\cc} \end{cases},
\label{vs}
\end{equation}
where $S_{\cc}$ is the critical gap, beyond which vehicles travel unrestricted by their leaders, $S_{\jj}$ is the jam gap, which  is the smallest distance that a vehicle maintains from their leaders (when the speed is zero), $w_{\bb}$ is the backward wave speed, which is the speed at which a platoon grows in traffic from standstill, and $v_{\ff}$ is the maximum speed when traveling unrestricted.  Locally, the backward wave speed is related to both the jam spacing and the reaction time as $w_b = \tau^{-1}S_{\jj}$.  Clearly, these parameters will vary from one vehicle to another but are bounded from both above and below.  They are, thus, represented by probability distributions with bounded domains.

To this end, we will assume (without loss of generality) that the parameters are drawn from generalized beta distributions. Let $A$ be any of the three parameters $v_{\ff}$, $w_{\bb}$, or $S_{\jj}$ with minimum and maximum values denoted by $A^{\min}$ and $A^{\max}$, and let $a_A$ and $b_A$ denote the shape parameters of the distribution of $A$. The probability density function (PDF) of $A$, denoted $p_A$, is
\begin{multline}
    p_{A}(x) = \frac{(A^{\max} - A^{\min})^{-a_A-b_A+2}}{B(a_A,b_A)}  (x - A^{\min})^{a_A-1} \\ \times (A^{\max}-x)^{b_A-1}   I_{[A^{\min},A^{\max}]}(x),  \label{beta1}
\end{multline}
where $B(\cdot,\cdot)$ is the beta function and $I_{[A^{\min},A^{\max}]}(x) = 1$ if $x \in [A^{\min},A^{\max}]$ and 0 otherwise.

The use of beta distribution has been justified in previous works by one of us ~\cite{jabari2014,jabari2018stochastic,zheng2018stochastic,jabari2020sparse}, the main advantage being
 the bounded support from above and below, and the flexibility in the shape of the distribution afforded by the Beta distribution. The heterogeneity as introduced here is nothing but a quenched disorder in each of the parameter. The $\tau$, because of its dependence on $w_{\bb}$ and $S_{\jj}$ is also a quenched disorder. The above choice of $\tau$ ensures there are no numerical or local instabilities.  However, we can expect string instability to appear and play a role in the collective dynamics of the system.
 
The system in the present work is a set of $N$ vehicles on a single lane track of 
length $L$ with periodic boundary conditions (PBC). Thus the (spatial) average density
on the road is $\bar \rho = NL^{-1}$. The vehicles are assumed to be spaced uniformly
initially (i.e., at $t=0$). The simulation consists of evolving the $N$ coupled delay differential 
equations with PBC. For the present work, we use the following set of 
parameters: $v_{\ff}^{\min} = 60$~km/h, $v_{\ff}^{\max} = 80$~km/h, $w_{\bb}^{\min}=30$~km/h, 
$w_{\bb}^{\max}=40$~km/h, $\rho_{\jj}^{\min} = 1/S_{\jj}^{\max} = 170$~veh/km, and 
$\rho_{\jj}^{\max} = 1/S_{\jj}^{\min} = 130$~veh/km. The time step is taken to be 
$\Delta t =0.5 \times 10^{-6}$~h. For the beta distributions of $v_{\ff}$ and $\rho_{\jj}$,
 we chose $a_{v_{\ff}}=2$ and $b_{v_{\ff}}=2$ (symmetric distribution), and for the beta distribution of $w_{\bb}$, we chose $a_{w_{\bb}}=2$ and $b_{w_{\bb}}=3$ (skewed).

The method used for solving the $N$ delay differential equations (DDEs) is similar to that discussed by  
Kesting et al.~\cite{Treiber2013}. The vehicles are placed at a uniform initial gap i.e., $s_i(0) = LN^{-1}$ and each vehicle
 is initiated with its free-flow speed i.e., $v_i(0)=v_{\ff,i}$ where $\{v_{\ff,i}\}_{i=1}^N$ are drawn from the beta distribution described above.  For numerical convenience, we approximate $\tau_i~(\gg\Delta t)$ as $\tau_i \approx \mathrm{nint}(\tau_i/\Delta t)$ in units of $\Delta t$ where $\mathrm{nint}(x)$ is the nearest integer to $x$.  The basic difference between a DDE and the ordinary differential equation is that
 in a DDE the values of the coordinates for the past $\mathrm{nint}(\tau_i/\Delta t)$ time  steps have to be memorized for each vehicle. For simplicity, we assume that $s_i(t - \tau_i) = LN^{-1}$ for all $t=0,\hdots,\tau_i$ and all $i$, which implies that $v_i(t) = v_i(s(t-\tau_i)) = 0$ for all $t < \tau_i$. The update equations for each vehicle at each time step are as follows:
 \begin{equation}
    s_i(t+\Delta t) = s_i(t) + {\Delta t}(v_{i-1}(t)-v_{i}(t)),
 \end{equation}
 where
 \begin{equation}
     v_i(t) = V(s_i(t-\tau_i))
 \end{equation}
 as given in Eq.~\eqref{vs}. 
 After each step, the memory of speeds and gaps is updated to include the past $\tau_i$
 values.
 
 \subsection{Flow instability} 
The delayed reaction by drivers as modeled by the coupled DDEs in Eq.~\eqref{EOM} introduces oscillations in the gaps between vehicles and in their speeds. This is considered as a form of instability in traffic flow, similar to instabilities in fluid dynamics. Instabilities in traffic flow are broadly classified as local and string instabilities. A system of vehicles is locally unstable if the gap and speed fluctuations of each vehicle do not decay with time. A string instability, as the name suggests, is that in which a perturbation in the gap (and the speed) travels upstream in a manner similar to a traveling wave in a string. If the conditions in the system are such that the amplitude of the perturbation increases as it travels
upstream, the perturbation eventually transforms into a jam, within which the vehicle(s) come so close to each other that they either move very slowly or halt momentarily. The jam front thus formed continues to move upstream forming what is called a \emph{stop-go wave}.
In the appendix, we illustrate the role played by the reaction time in inducing oscillations in the gap (and hence in the speed)  by deriving an approximate analytical expression for the gap and speed for the case of a follower equilibrating their speed to that of a slow moving leader.  Here, we explain in simple terms, the way in which a perturbation in the speed of one vehicle gets amplified into a stop-go wave as it spreads to the vehicles upstream. We refer the reader to Chapter 15 in Treiber and Kesting's book \cite{Treiber2013} for a  more detailed discussion on instabilities in traffic flow.

\begin{figure}
	\resizebox{0.49\textwidth}{!}{%
\includegraphics[angle=0]{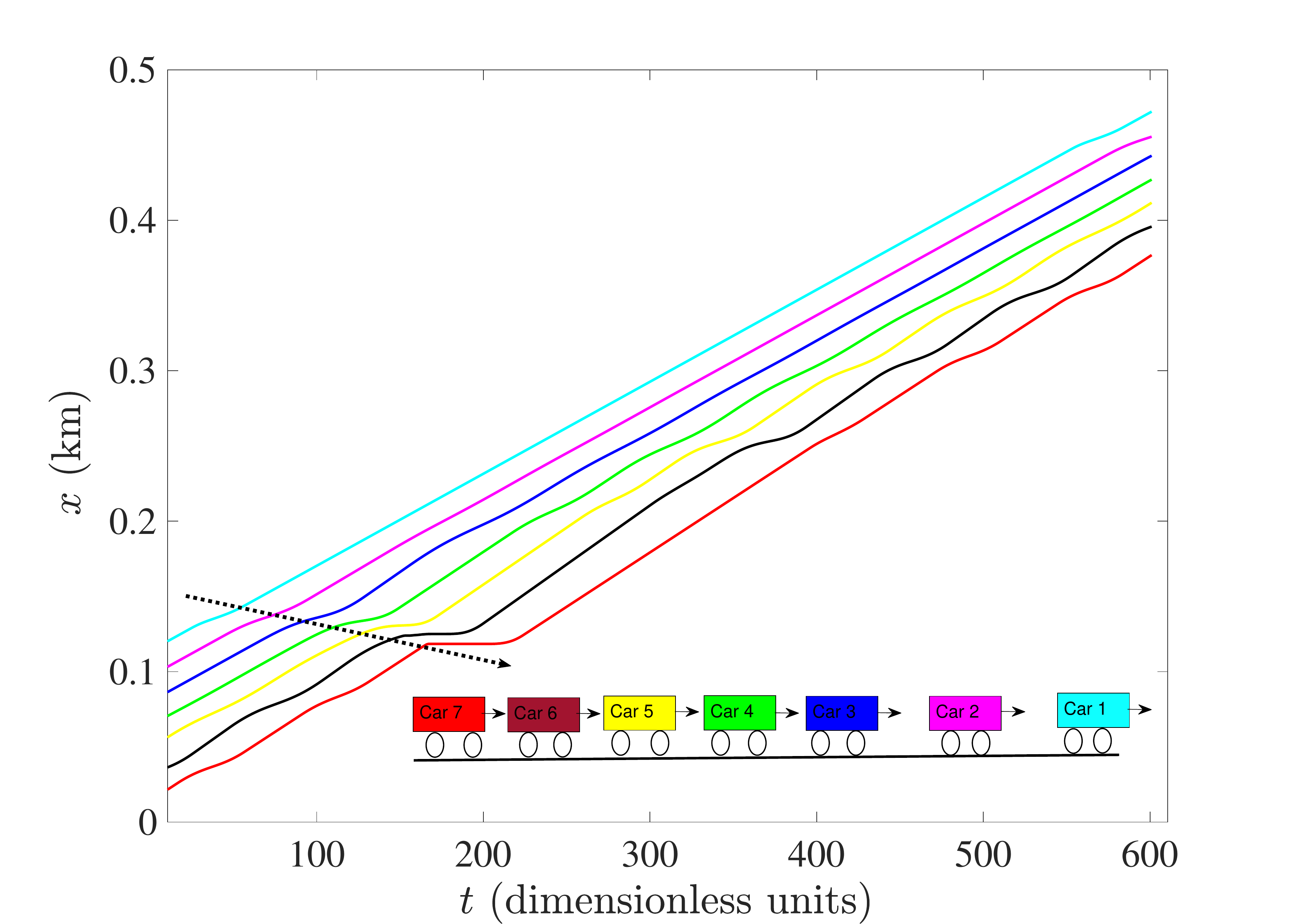}}

\resizebox{0.49\textwidth}{!}{%
\includegraphics[angle=0]{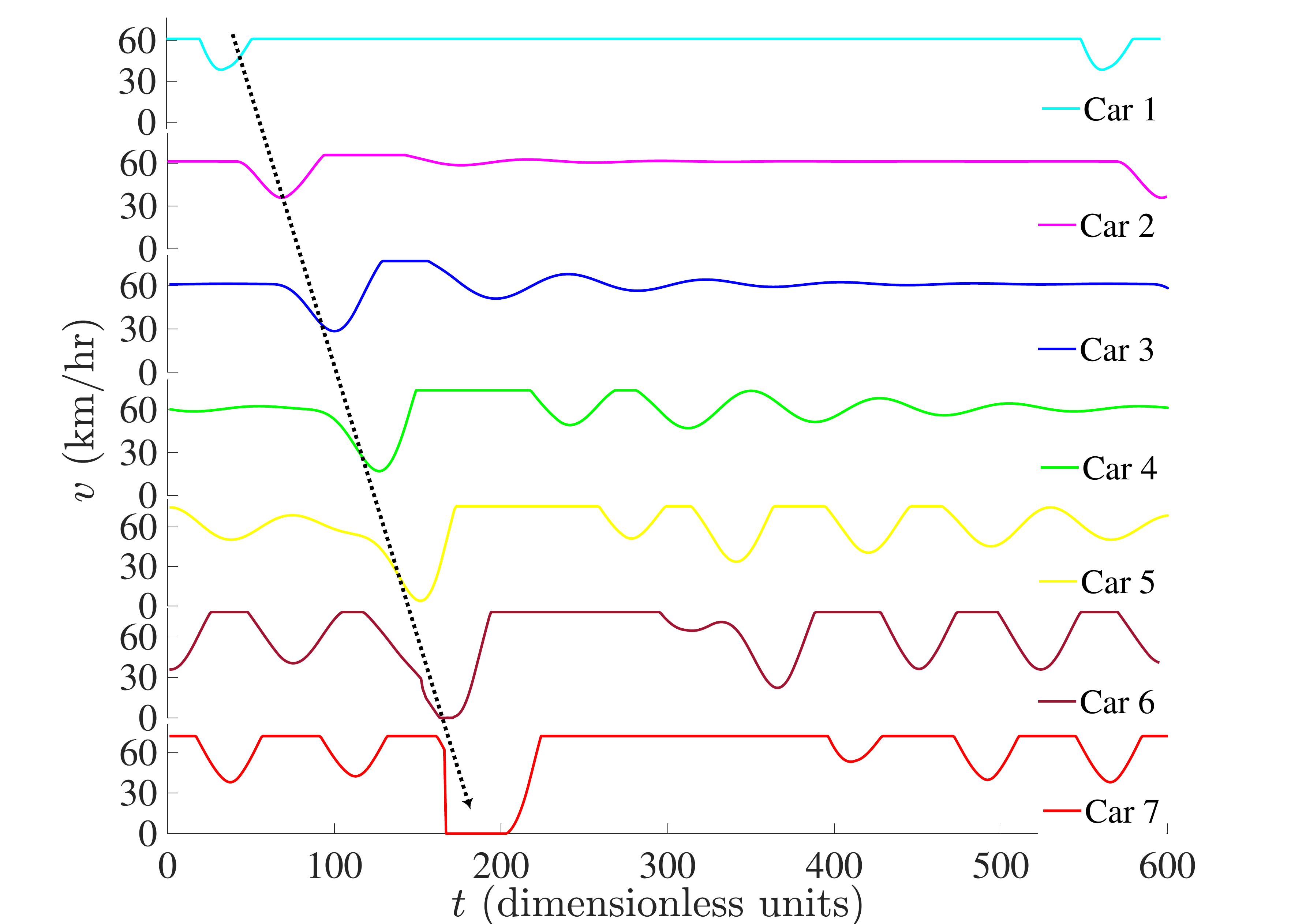}
}
\caption{\label{fig:instx} (Top) Position ($x$) versus time ($t$) plot illustrating the formation of a stop-go wave. The vehicles move from left to right as depicted in the cartoon. The curves corresponds to the cars with matching colors.
(Bottom) Speed ($v$) versus time ($t$) plot for the vehicles in the top figure. Dotted arrows are guide to the eye.  A small trough in the speed of vehicle 1  induces a larger trough in vehicle 2 and so on. As this reaches vehicle 7, the 
perturbation develops into a stop-go wave. The damped oscillations occur as the vehicle equalizes its speed with that of its leader.}
\end{figure}

The speed versus time plot for a platoon of seven cars with the $i$th car following car $i-1$ is shown in Fig.~\ref{fig:instx}.  The simulations are done using the  parameter settings described in the previous section. For that choice of parameters, the system exhibits only string instability. Cars 1 and 2 initially travel at the same speed. Car 2 reacts to the slight slow down of car 1 (see the trough near top-left corner in the figure) with a delay equal to its reaction time. Because of the delay, the gap ahead of car 2 shrinks by more than the steady state gap required for its initial speed. As a result, car 2 reduces its speed to a value less than that of car 1 to maintain a safe distance. Thus the depth of the trough in its speed plot is more pronounced than that of car 1. Similarly, car 2 perceives the increased speed of car 1 with a delay. The culmination of these two maneuvers (excessive slowing and delay) is a large gap ahead of car 2.  As a result, car 2 increases its speed to more than that of car 1 briefly. However, as the gap between them decreases, car 2 again starts equilibrating its speed with that of car 1 and the delay time again causes over-shooting and under-shooting of the speed of  car 2, which results in the oscillations seen in Fig.~\ref{fig:instx}. This effect cascades as it spreads upstream and car 7 comes so close to car 6 at some point that it has to stop and thus begins a stop-go wave, which continues to move upstream with cars following car 7 completely stopping momentarily and starting again to move. As can be seen from the figure, the oscillations in each car tend to decay. However, in some cases, a vehicle experiences another perturbation from its leading vehicle before the oscillations from previous perturbation totally decay, because of which the vehicle would experience persistent oscillations. Thus, the heterogeneity of the system and the delay in response due to reaction-time make the system extremely complex and are basically the origins for the oscillations and the stop-go waves.

\subsection{Relation to  three-phase theory}
Before we conclude our brief introduction of car-following theory, we briefly present \emph{three-phase theory} and its relation to the present context.  Newell's model, as presented above, is considered to be a \emph{two-phase} model; the two phases being the free-flow phase ($\mathsf{F}$) and the congested phase, which we will refer to as the \emph{jam} phase ($\mathsf{J}$) to be consistent with the nomenclature used in three-phase theory.  The two phases can be seen in the two parts of Newell's hypothecized speed relation in Eq.~\eqref{vs}.  Three-phase theory was developed by Kerner~\cite{Kerner1,Kerner2} who analyzed data from German autobahns and observed a third phase sandwiched in between $\mathsf{F}$ and $\mathsf{J}$, which he dubbed \emph{synchronized traffic} ($\mathsf{S}$).  The $\mathsf{S}$-phase is a congested phase with no ``wide-moving jams'' (stop-go waves). The kinetics of the transition are similar to a nucleation process; the $\mathsf{S}$-phase forms (or nucleates) near a traffic bottleneck (e.g., off-ramps and on-ramps on a highway) and if the conditions are favorable, the size of the `nucleus' keeps growing in the upstream traffic direction (from the bottleneck). However, as the size of the region of the $\mathsf{S}$-phase increases, the flow becomes unstable and stop-go waves emerge. The formation of the stop-go waves is called an $\mathsf{S} \rightarrow \mathsf{J}$ transition. While the above phenomenon occurs during a transient state, the system may reach a non-equilibrium stationary state with the same flow pattern.
Thus, one finds a $\mathsf{F}$-phase ahead of the bottleneck and as one goes upstream, first an $\mathsf{F} \rightarrow \mathsf{S}$ transition is observed near the bottleneck and then an $\mathsf{S} \rightarrow \mathsf{J}$ transition is observed at a point further upstream. 

Proponents of the three-phase theory have sharply criticized two-phase theories on the grounds that they all fail to capture the nucleation process described above (the most recent criticism appeared in Appendix~A in \cite{kerner2021effect}). This criticism has been extensively debated in the traffic flow literature. A number of works showed that the empirical observations mentioned above can be simulated using standard car-following models with a proper choice of parameters. See, for example \cite{Schonhof2007tpt,Schonhof2009tpt,Treiber2010tpt}.  For example, in the simulations performed in this study depicted in Fig.~\ref{fig:xVstrt}, the dark patterns in the plot that form somewhere in the middle of the platoon and move upstream are the stop-go waves (phase $\mathsf{J}$). The leader of the platoon is the slowest vehicle in the system. It experiences free-flow but being the slowest vehicle in the system it plays the role of a moving bottleneck for the faster vehicles behind it.  It can be seen from the figure that the immediate followers of the slowest vehicle experience 
synchronized flow i.e., the $\mathsf{S}$-phase. We can also see from the figure that the $\mathsf{S}$-phase doesn't spread indefinitely in space; it becomes unstable as it moves upstream where 
stop-go waves appear in the system. Therefore, starting from the leader of the platoon and moving upstream, 
one sees a  $\mathsf{F} \rightarrow \mathsf{S}$ transition near the leader (the bottleneck in our case) and a $\mathsf{S} \rightarrow \mathsf{J}$ transition at a point further upstream.  Fig.~\ref{fig:xVstrt} may be compared to Figure 1.3 in Ref.~\cite{Kerner2}.  We note that the presence of quenched disorders in the parameters of our two-phase model is a unique feature in our model.  Further, it was argued by the defenders of the two-phase theories that the classification of the $\mathsf{S}$-phase and the $\mathsf{J}$-phase separately and the introduction of $\mathsf{F} \rightarrow \mathsf{S}$ \emph{and} $\mathsf{S} \rightarrow \mathsf{J}$ transitions was just qualitative with a different interpretation being possible.  It was also argued that the observed pattern of transitions, $\mathsf{F} \rightarrow \mathsf{S}$ followed by $\mathsf{S} \rightarrow \mathsf{J}$, does not always occur in real traffic. 

The purpose of the present work is to investigate the impact of a quenched disorder in the reaction times.  We make no claims of addressing the ongoing two-phase and/or three-phase debate.  While we elected to use a two-phase model for simplicity of exposition, we believe that quenched disorders, particularly in the reaction times, can have profound impacts on emergent phenomena in traffic independent of whether a two-phase or three-phase theory is used.

\section{Approach to stationary state} 

\begin{figure}
	\resizebox{0.49\textwidth}{!}{%
\includegraphics[]{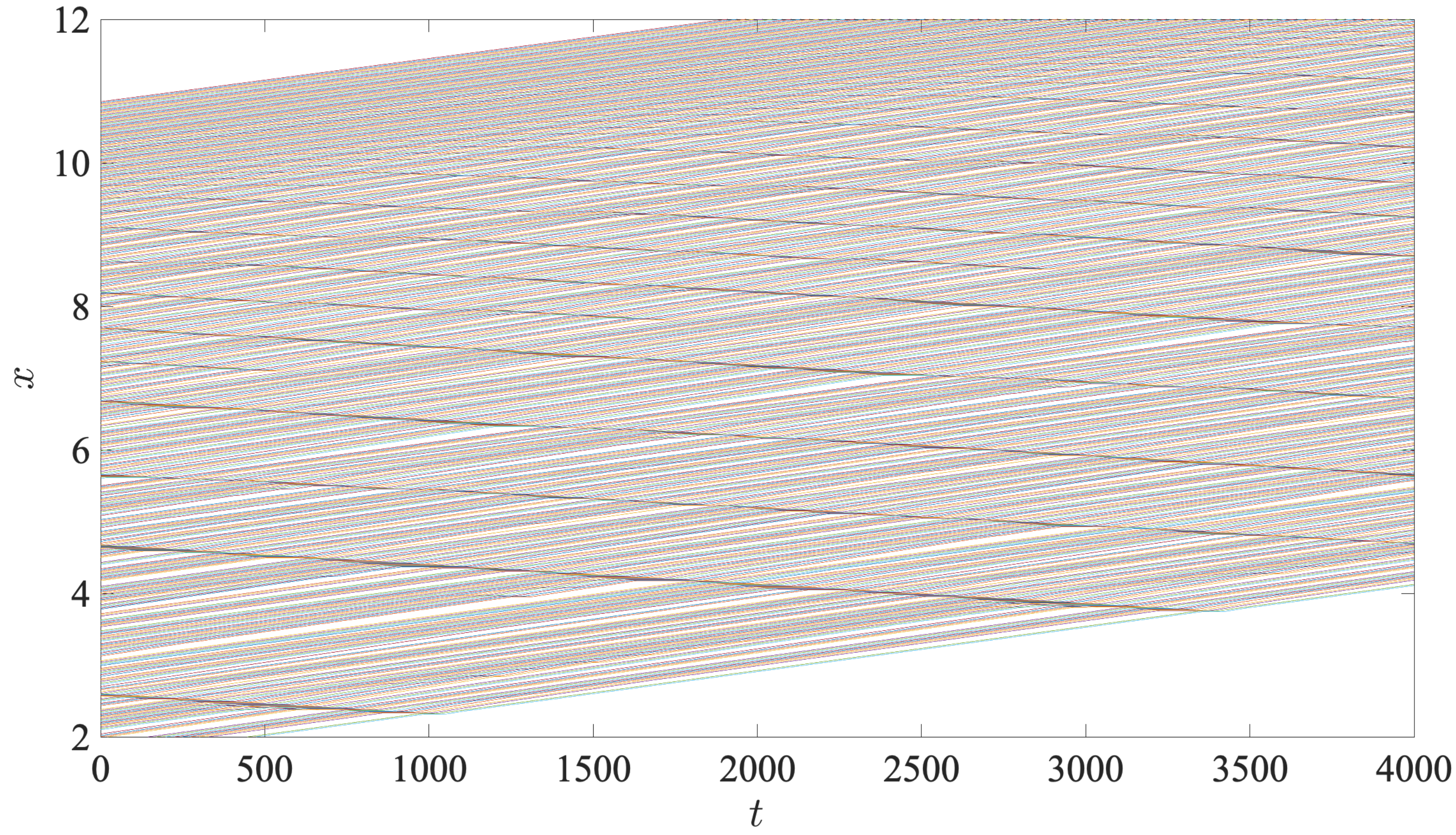}
}
\caption{\label{fig:xVstrt} A plot of typical positions of the $N$ vehicles versus time after the stationary state is reached when reaction time is included. Each line corresponds to a vehicle. The vehicles at the front of the platoon move smoothly without any stop-go waves. Some small oscillations generated in the vehicles at the front develop into stop-go waves as they move upstream.  The stop-go waves (dark patterns moving upstream in the figure) get dissipated at the end of the platoon.}
\end{figure}

We observed that the  flow instability induced by the chosen values of $\tau$ doesn't hinder the formation of a single platoon at low densities; see Fig.~\ref{fig:xVstrt}.
As explained above, some small perturbations in the gaps of the vehicles in platoon get amplified as they go upstream 
of the platoon and form stop-go waves. However,  the strength of the instability doesn't grow indefinitely. We observed that the stop-go waves may get totally dissolved or the number of vehicles participating in the stop-go waves keep fluctuating. This may happen because of various factors, e.g., a large gap between the leader and the follower or an agile follower with small reaction time and small critical gap. Thus the phase-ordering due to the quenched disorder in speed wins over the instability due to the reaction time when the single platoon forms.

As explained in the introduction, the NS model with a quenched disorder in hopping rates is understood to be belonging
to a general dynamical universality class with $z = (n+2)/(n+1)$ where $n$ is the exponent of the distribution of the quenched disorder. Ben Naim et al.~\cite{Ben1994} analytically derived the same $z$ value for the case of one dimensional ballistic aggregation which may be related to a one dimensional car following model with a quenched disorder in free-flow speed. In this case, $n$ was the exponent of the  distribution for $v_{\ff}$ close to $v_{\ff}^{\min}$. Thus, for the beta distribution for $v_{\ff}$ used here, $n = a_{v_{\ff}} -1$. In our previous work, we simulated Newell's model with quenched disorders in the $v_{\ff}$, $S_{\jj}$ and $w_{\bb}$ with zero reaction time and obtained the $z$ numerically and using finite size scaling, which matched with that of Ben Naim et al. and of the NS model. In the present case, where we include a reaction time for each driver, it is not clear whether the system belongs to the same universality class as the collective effects due to the string instability induced by the reaction time oppose the formation of a platoon and thus we determine it below.  

\begin{figure}
	\resizebox{0.49\textwidth}{!}{%
\includegraphics[]{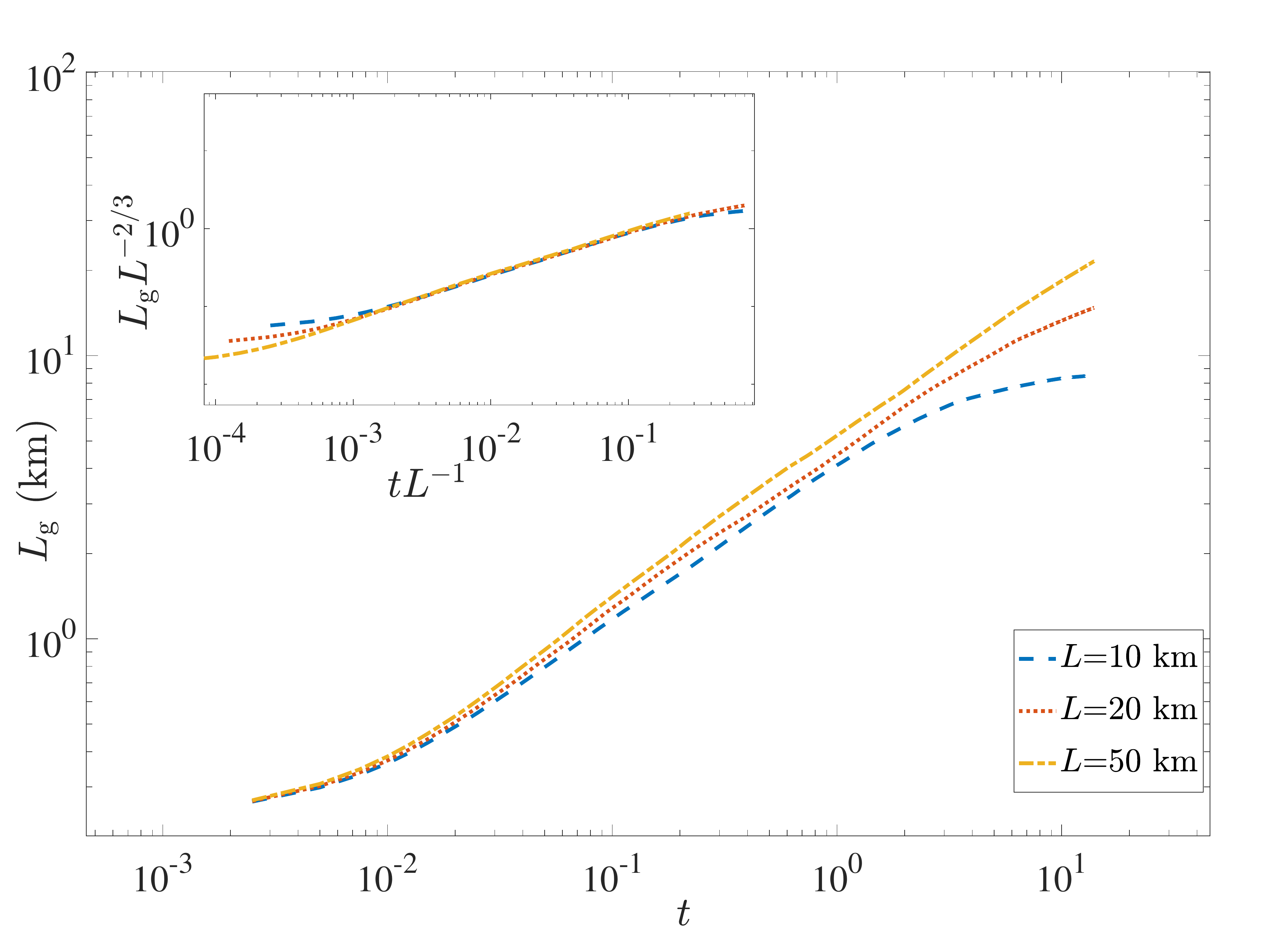}
}
\caption{\label{fig:gaprt} Gap size as a function of time. Scaled gap length versus scaled time is shown in the inset. Collapse of curves upon the scaling implies $L_{\gx} \sim t^{2/3}$. }
\end{figure}

\begin{figure}
	\resizebox{0.49\textwidth}{!}{%
\includegraphics{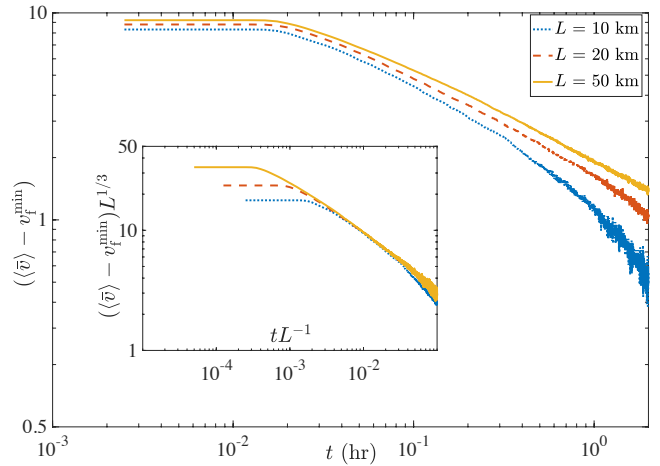}
}
\caption{\label{fig:vrt}  Average speed as a function of time for various track lengths. Scaled average relative speed is shown in the inset. Collapse of curves with scaling implies that $\langle \bar v \rangle \sim t^{-1/3}$.}
\end{figure}

The fluctuations in the gaps between the vehicles 
made it extremely difficult to identify the size (or length)
of the platoon after the string instability grows strong. Thus, we determined the size of the largest gap ($L_{\gx}$) in the system as a function of time instead of the average platoon size. As the growth of $L_{\gx}$ implies the growth of the platoon, both should follow the same power-law. 
The averages have been calculated typically over few tens of independent simulations each done with a different random seed for the quenched disorders. From Fig.~\ref{fig:gaprt}, it can be seen that a typical gap grows as a power-law. Clearly, the finite size effects are quite dominant. A finite size scaling form is also depicted in the inset of Fig.~\ref{fig:gaprt} and confirms the power-law exponent to be $z = 3/2$ which matches with that of the NS model with quenched disorder. Similarly, the power-law exponents for the average relative speed $\langle v - v_{\ff}^{\min}\rangle$
turns out to be the same as the previous case i.e., $\alpha_s = -1/3$ which also matches with that derived by Ben Naim et al. A typical power-law decay of $\langle v - v_{\ff}^{\min}\rangle$ and its finite size scaling form are shown in Fig.~\ref{fig:vrt}. Thus we see that while the string instability complicates the platoon dynamics, it doesn't alter the dynamical exponent for platoon formation which may be a confirmation of  the
kinematic wave criterion as argued by Tripathi and Barma~\cite{Tripathy1997,Barma2006}. 

\section{Characterization of the giant platoon} 
As seen in Fig.~\ref{fig:xVstrt}, at low densities, the system which was initially spatially homogeneous drifts into a stationary state where all the vehicles segregate into a single
platoon with the slowest vehicle leading it and a large system size dependent
gap ahead of the slowest vehicle.
The difference between the present case with reaction time and the case with no reaction
time is the presence of stop-go waves. 
For a finite system, the stop-go waves are not observed at
very low densities. However, when the system size is increased keeping the density constant, the stop-go waves
emerge. Therefore we note that the stop-go waves exist at all non-zero densities in the large system limit (or in the thermodynamic limit). 
 
The stationary gap distribution $p(s)$ helps in characterizing the state of the system. We calculated $p(s)$ using the binning method. To coarse-grain the fluctuations at small time scales, we took averages over sufficiently long time which is typically few tens of thousands of steps after the stationary state is reached. Although the system is expected to be ergodic (having a unique stationary state) and self-averaging in the thermodynamic limit, to avoid any initial state dependence because of finite system size and to smooth the fluctuations further, we do an ensemble average.  To perform calculations, we set a bin size of $\Delta s$ and compute the probability density as
  \begin{equation}
  	p(s) = \frac{1}{E}\frac{1}{T}\sum_{e=1}^E\sum_{t=1}^T \frac{N_t^e(s)}{N \Delta s}
  \end{equation}
  where $E$ is the number of ensemble copies, $T$ is the number of time steps over which averaging is done and $N_t^e(s)$ is the number of vehicles in ensemble $e$ during time step $t$ that have a gap between $s$ and $s+\Delta s$. We determined the distributions for track lengths $L=5,10,20,50$ and $100$ ~kms. For $L=5$~km, the ensemble averaging is done over $200$ copies while for $L=100$~km,  averaging is done over $24$ copies. The number of copies for remaining lengths are between $200$ and $24$. The copies are chosen as a compromise between the smoothness of the obtained curves and the computational time.
   
\begin{figure}
	\resizebox{0.49\textwidth}{!}{%
    \includegraphics[]{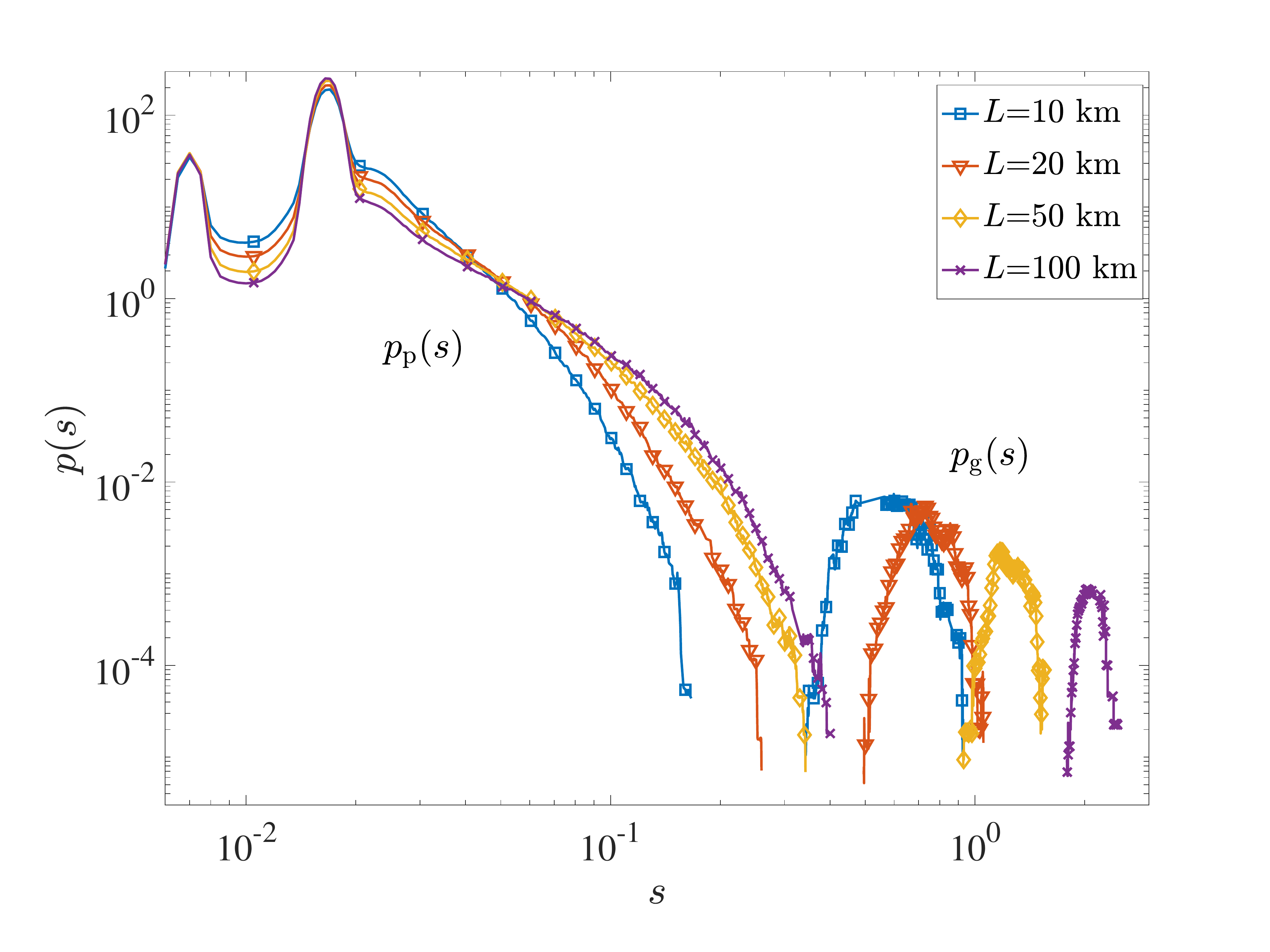}}
    \caption{Stationary state probability of gap $p(s)$; one part of it is the gap distribution behind the leader of the platoon $p_{\pp}$ (the left curve) and other part is the gap distribution ahead of the leader $p_{\gx}$ (the right curve). 
    }
    \label{fig:Psvssrt}
\end{figure}

A typical $p(s)$ at a low density where the platoon formation happens is depicted in Fig.~\ref{fig:Psvssrt} for various track lengths. $p(s)$ has two distinct components: the probability of gap behind the slowest vehicle ($p_{\pp}(s)$) and $p_{\gx}(s)$, which is the probability of gap ahead of the slowest vehicle. When there is no reaction time~\cite{Ramana2020}, we showed that $p_{\pp}(s)$, in the thermodynamic limit, is identical to $p_{\cc,l}(s)$ which is the critical gap distribution of the leader.  In the present case, $p_{\pp}(s)$ significantly differs from $p_{\cc,l}(s)$.  However, a peak in the distribution still appears at the gap where the $p_{\cc,l}(s)$ has a peak. In addition, another peak can be seen close to the jam-gap $S_{\jj}$. This peak appears as a result of stop-go waves. 
The broadening of the $p_{\pp}(s)$ on the right side is because of the gaps of various sizes that get created due to the stop-go waves. It can be seen from the Fig.~\ref{fig:Psvssrt} that the upper bound of the distribution increased with
the increase in track length but tends towards a converged value. Thus it becomes a fat-tailed distribution in the thermodynamic limit. The $p_{\pp}$ part is found to be independent of density until the phase transition point is reached. The  $p_{\gx}(s)$ can also be seen in the figure. The distribution is much broader than the case with no reaction time. Thus the $p(s)$ is dominated by the flow instability and the stop-go waves. As density is increased, $p_{\gx}(s)$ shifts closer to $p_{\pp}(s)$ and merges with it on approaching the phase transition point.
\section{The dynamical phase-transition}
As one goes from low density to high density, a density point is reached after which the giant platoon doesn't form in the stationary state. In simple terms, one may anticipate the transition to happen when the head of the platoon starts interacting with the tail of the platoon.  For the case with no reaction time, we showed that the transition is always of first order following the conjecture by Krug et al.~\cite{Krug1996} that there is no divergence in the variance of the stationary gap distribution at the transition point for any choice of parameters describing the quenched disorders. We also showed that, in the thermodynamic limit, the density ($\rho_{\cc}$) at which the transition happens is actually the reciprocal of the expectation value of the gap distribution $p_{\pp}(s)$ behind the slowest vehicle. In the present case, when a reaction time for each driver is considered, we observed that this is still approximately  valid i.e.,  
\begin{equation}
   \frac{1}{\rho_{\cc}} \approx \langle s \rangle_{\pp} \approx \int \dd s p_{\pp}(s) s,  \label{scrt}
\end{equation}
which becomes exact in the thermodynamic limit. We noted from numerical calculations that the $\rho_{\cc}$ in the present case is less than that with zero reaction time. Because of the finite size effects in $p_{\pp}$, the calculated the $\rho_{\cc}$ also has a length dependence. However, we found that the tail of the distribution has much lesser weight and therefore the expectation values calculated for track lengths of $50$~km and $100$~km were pretty close. To determine the transition density in the thermodynamic limit ($\rho_{\cc_{\infty}}$), we fitted the the numerically observed $\rho_{\cc}$ for various lengths to the below form:
\begin{equation}
	\rho_{\cc}(L) = \rho_{\cc_{\infty}} + \frac{B}{L^\nu} \label{rhocfit}
\end{equation}
and obtained $\rho_{\cc_{\infty}} \approx 48.71$, $B \approx 13.22$ and $\nu \approx 1.03$ for the present case.

\begin{figure}
	\resizebox{0.49\textwidth}{!}{%
     \includegraphics[]{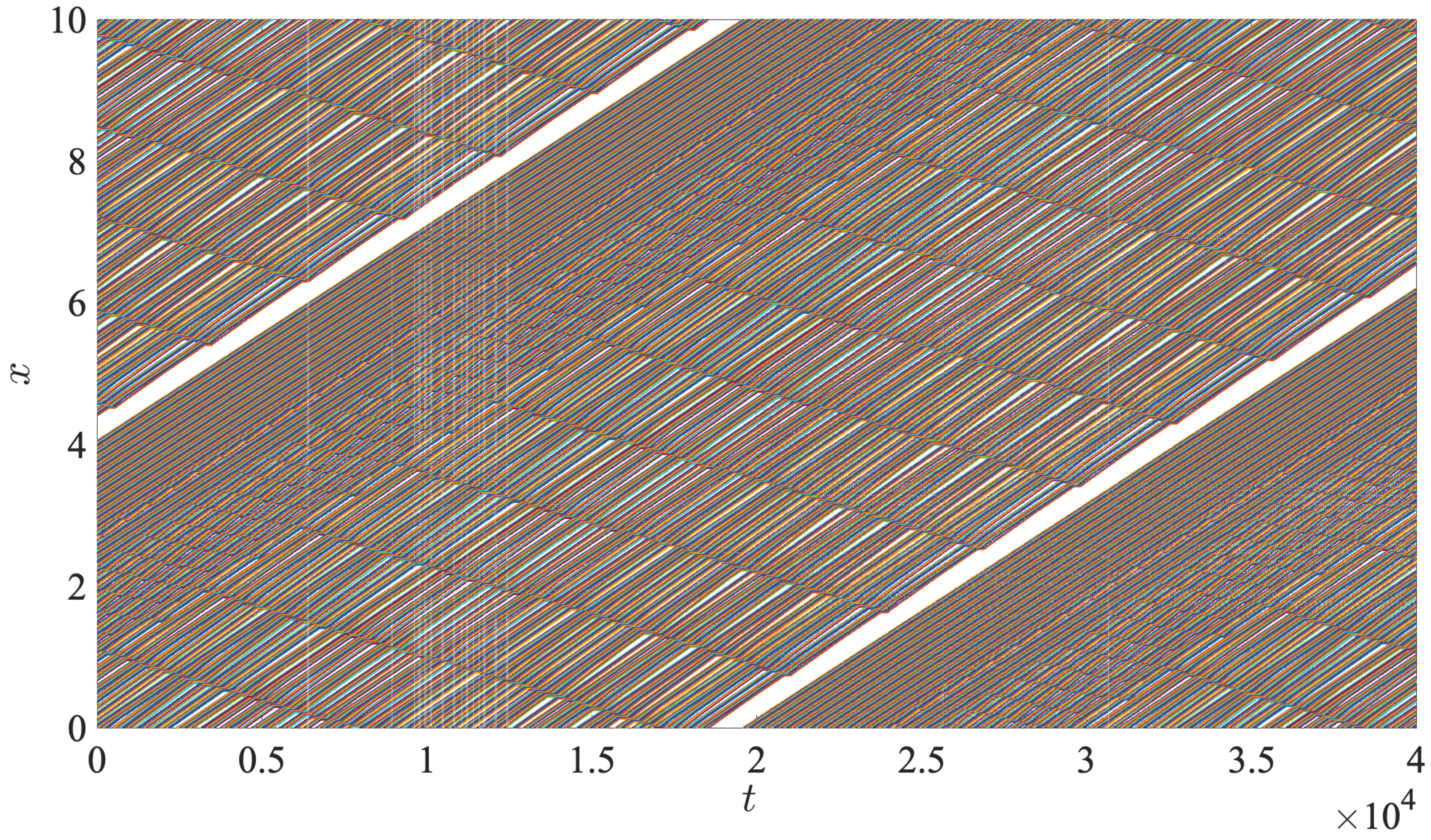}}
     
     \resizebox{0.49\textwidth}{!}{%
     \includegraphics[]{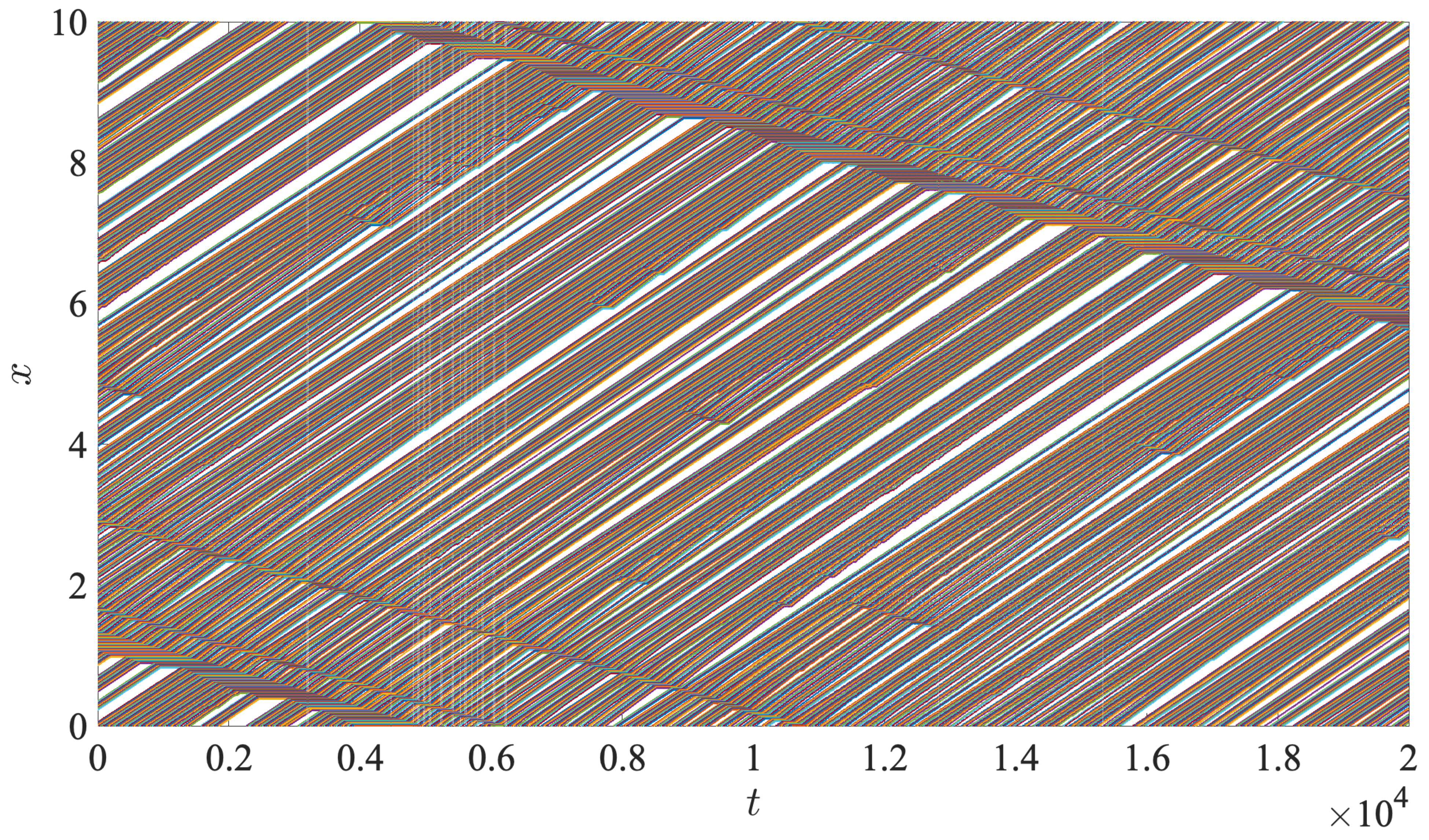}}
     \caption{Positions of the $N$ vehicles versus time on a $10$~km track. For this case, $\rho_{\cc}$ lies between 49 and 50~veh/km. Top plot is for $\rho =49$~veh/km, which is below $\rho_{\cc}$. Bottom plot is for $\rho =50$~veh/km, which is above $\rho_{\cc}$.}
     \label{fig:xvst-rt}
 \end{figure}

 To illustrate the phenomenon, we
 plot  in Fig.~\ref{fig:xvst-rt} the positions of the $N$ vehicles versus time just below and above the predicted $\rho_{\cc}$. For ease of visualization, the $L=10$~km case is depicted in the plot.
 The following points may be noted by observing the plots. Below $\rho_{\cc}$, the collective effect of
platoon formation is dominant and a single giant platoon is formed. Some stop-go waves generated in the middle of the platoon are stable and travel upstream to the end of it where they get dissipated. Above $\rho_{\cc}$, the collective effect due to flow instability becomes dominant and the formation of the giant platoon is hindered by strong stop-go waves, which move uninterrupted upstream all around the ring. As a result, there is a continuous process of formation and destruction of platoons.  Therefore, as the density is increased from a low value to $\rho_{\cc}$, the transition occurs when the stop-go waves start to dominate over the platoon formation due to the quenched disorder in $v_{\ff}$. Thus, the phenomenon happening at the phase transition is more complicated than mere interaction of the head of the platoon with its tail.

To characterize the transition, we determined some physical quantities in the stationary state over a range of densities above and below $\rho_{\cc}$ for various track lengths, which we analyze below. 
\begin{figure}
	\resizebox{0.49\textwidth}{!}{%
     \includegraphics[]{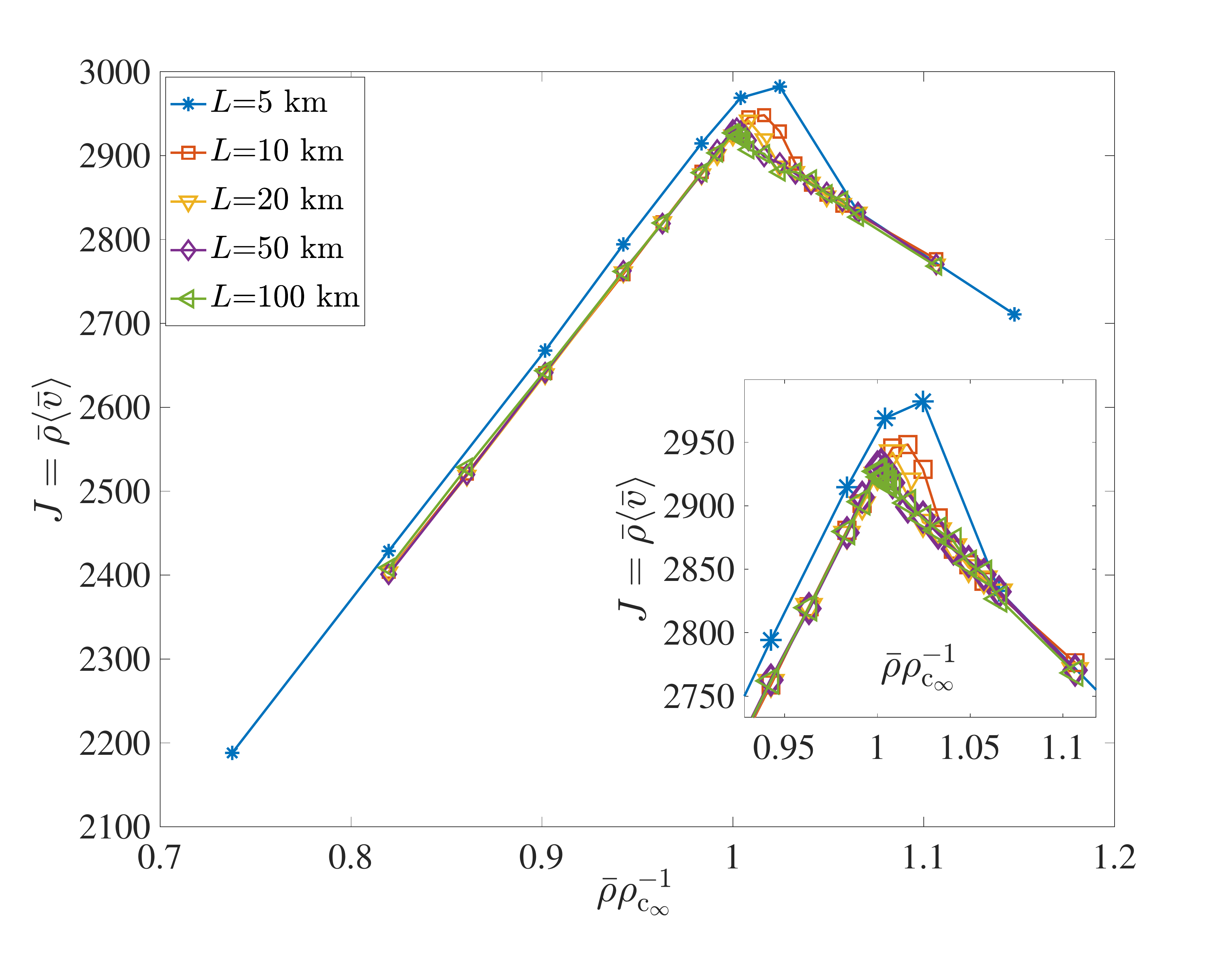}}
     \caption{Flow versus average density when reaction time is included}
     \label{fig:jdrt}
 \end{figure}
The flow-density diagram is plotted for various track lengths in Fig.~\ref{fig:jdrt}.  The finite size effects appearing in the diagram (see inset) are similar to those observed by Balouchi and Browne~\cite{Balouchi2016} and become negligible for track length of $100$~km and the diagram converges to a triangular shape.  It can be observed that the free-flow to congestion transition also happens at the density predicted by Eq.~\eqref{scrt}.
 
  \begin{figure}
  	\resizebox{0.49\textwidth}{!}{%
     \includegraphics[]{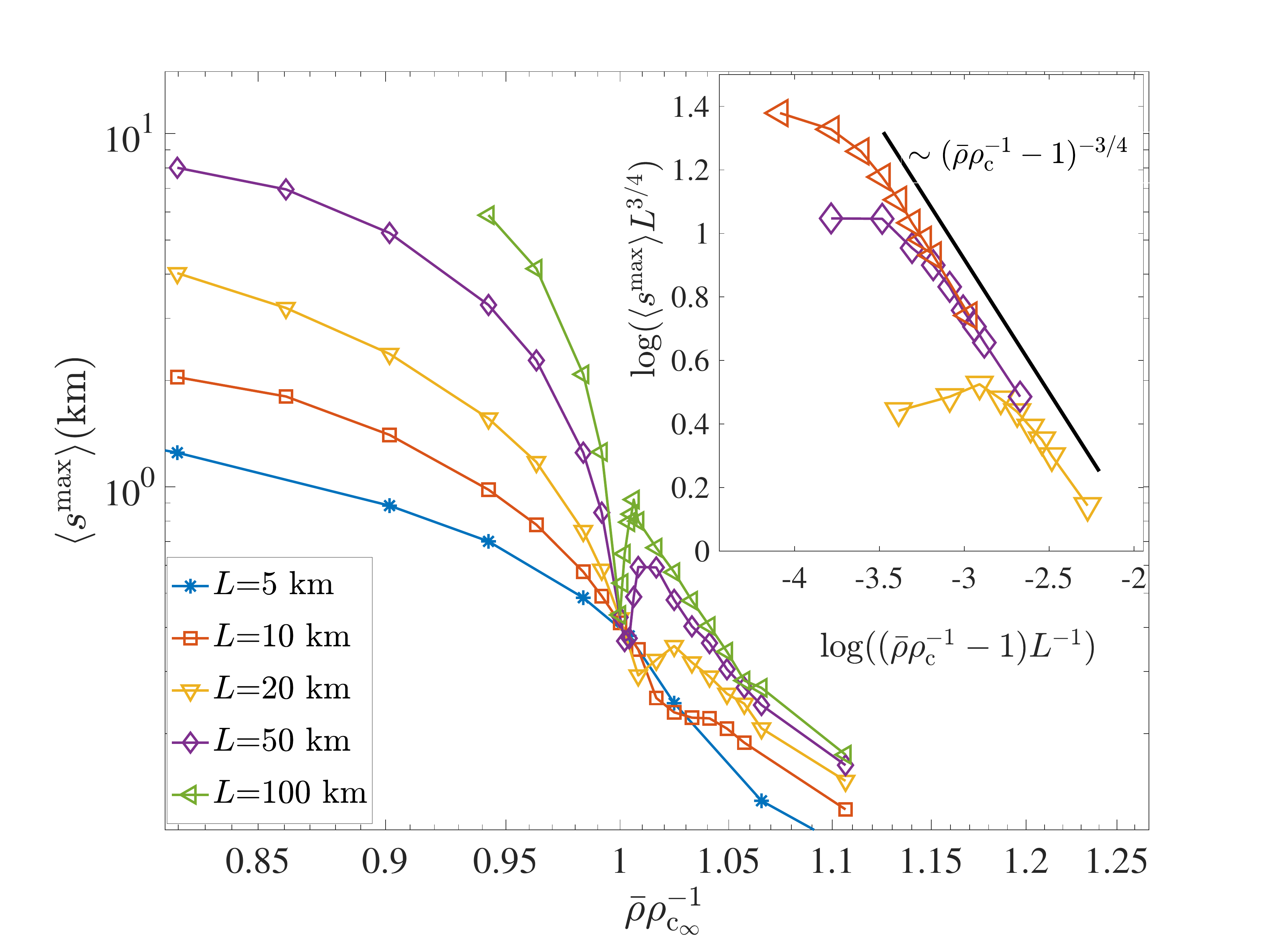}}
     \caption{Maximum gap versus density when the reaction time is included. Inset: Plot depicting the scaling $\langle s^{\max} \rangle \sim \big(\bar \rho \rho_{\cc}^{-1} - 1\big)^{-\gamma}$ with $\gamma=3/4$.}
     \label{fig:gap-d-rt}
 \end{figure}
 
  \begin{figure}
  	\resizebox{0.49\textwidth}{!}{%
     \includegraphics[]{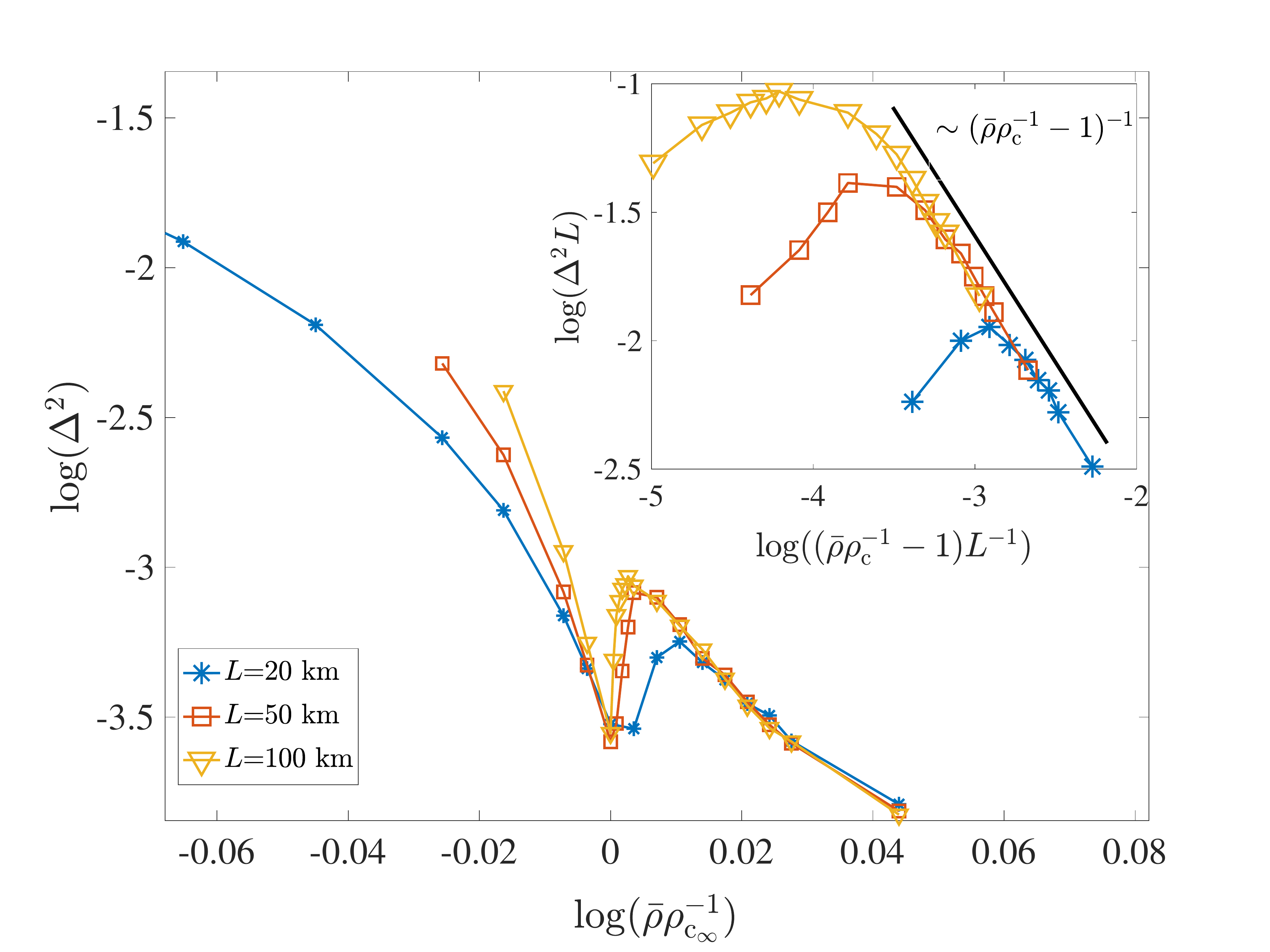}}
     \caption{ Variance of the gap distribution as function of density. Inset: Plot depicting the scaling $ \Delta^2  \sim (\bar \rho \rho_{\cc}^{-1} - 1)^{-\eta}$ with $\eta=1$}
     \label{fig:var-d-rt}
 \end{figure}

The average maximum gap $\langle s^{\max} \rangle$ observed in the system after the stationary state is reached is plotted against average density $\bar \rho = NL^{-1}$ in Fig.~\ref{fig:gap-d-rt} (scaled by $\rho_{\cc_{\infty}}$). For small values of $\bar\rho \rho^{-1}_{\cc_{\infty}}$, $\langle s^{\max} \rangle$ is proportional to $L$.  As $\bar\rho $  approaches $\rho_{\cc}$ from below, $\langle s^{\max} \rangle$ decreases and reaches a minimum value at $\rho_{\cc}$.  As $\bar\rho $ approaches $\rho_{\cc}$ from above, $\langle s^{\max} \rangle$ increases and resembles a power-law. A numerical fit of $\langle s^{\max} \rangle$ to $(\bar \rho \rho_{\cc}^{-1} - 1)^{ -\gamma}$ using data for $100$ km road gave $\gamma \approx 0.8$. Assuming the below finite-size scaling form 
\begin{equation}
	\langle s^{\max} \rangle \simeq L^{-\gamma \nu}f(XL^{-\nu})
\end{equation}
where $X = (\bar \rho \rho_{\cc}^{-1} - 1)$ and $f(X) \sim X^{-\gamma}$, the curves for various lengths collapsed when $\nu=1$ and $\gamma=3/4$. The value of $\nu$ agrees with that determined from Eq.~\eqref{rhocfit} and the value of $\gamma$ agrees well with the above numerically determined value. The gap variance
\begin{equation}
	\Delta^2 = \big\langle (s - \langle s \rangle)^2 \big\rangle = \int \dd s p(s) s^2 - \Big(\int \dd s p(s) s\Big)^2 \label{delta2}
\end{equation}
is plotted as a function of $\bar \rho$ in Fig.~\ref{fig:var-d-rt} for various $L$. Similar to $\langle s^{\max} \rangle$, $\Delta^2$ is proportional to $L$ for a given $\bar \rho$ and tends to a minimum size-independent value as $\rho_{\cc}$ is approached from below. As $\rho_{\cc}$ is approached from above, $\Delta^2$ increases as shown
in Fig.~\ref{fig:var-d-rt}. A numerical fit of $\Delta^2$ to $(\bar \rho \rho_{\cc}^{-1} - 1)^{-\eta}$ using data for 100 km road gave $\eta \approx 1.05$. We assumed the finite-size scaling form
\begin{equation}
	\Delta^2 \simeq L^{-\eta \nu}f(XL^{-\nu})
\end{equation}
where  $f(X) \sim X^{-\eta}$ and found that the curves for different lengths collapse in the power-law regime as shown in the inset of Fig.~\ref{fig:var-d-rt} when $\nu=1$ and $\eta=1$ . 

Thus, in both $\langle s^{\max} \rangle$ and $\Delta^2$, we notice 
power-law behavior as the $\rho_{\cc}$ is approached from above. There is lack of data for much larger track lengths to clearly visualize the power-law over an extended domain in the log-plots as simulations become highly computationally costly because of the power-law relaxation time required to reach the stationary state. 
However, we believe that the finite-size scaling clearly revealed the exponents $\gamma$ and $\eta$.  As a consequence of the different behaviors of $\langle s^{\max} \rangle$ and $\Delta^2$ as $\rho_{\cc}$ is approached from above  and  below, a kink can be observed in these quantities in the neighborhood of $\rho_{\cc}$ whose sharpness increases with an increase in track length $L$. We attribute the kinks observed in Figs.~\ref{fig:gap-d-rt} and \ref{fig:var-d-rt}, when $L$ gets large, to the emergence of vehicle clusters \emph{in between the stop-go waves}.  When $\rho$ is only slightly larger than $\rho_{\cc}$, the stop-go waves start to become pronounced but occur infrequently.  For large $L$, the distances separating the stop-go waves also get large, allowing for large $\langle s^{\max} \rangle$ to emerge in between the stop-go waves.  In this regime ($\rho \rightarrow \rho_{\cc}^+$) large $L$ also allows for higher variability in the cluster sizes to emerge near $\rho_{\cc}$, hence the increase in $\Delta^2$ observed above.  These kinks become less pronounced (and start to vanish) when $L$ gets small.  In the thermodynamic limit ($L \rightarrow \infty$) we expect the kinks to become infinite discontinuities.

\begin{figure}
	\resizebox{0.49\textwidth}{!}{%
		\includegraphics[]{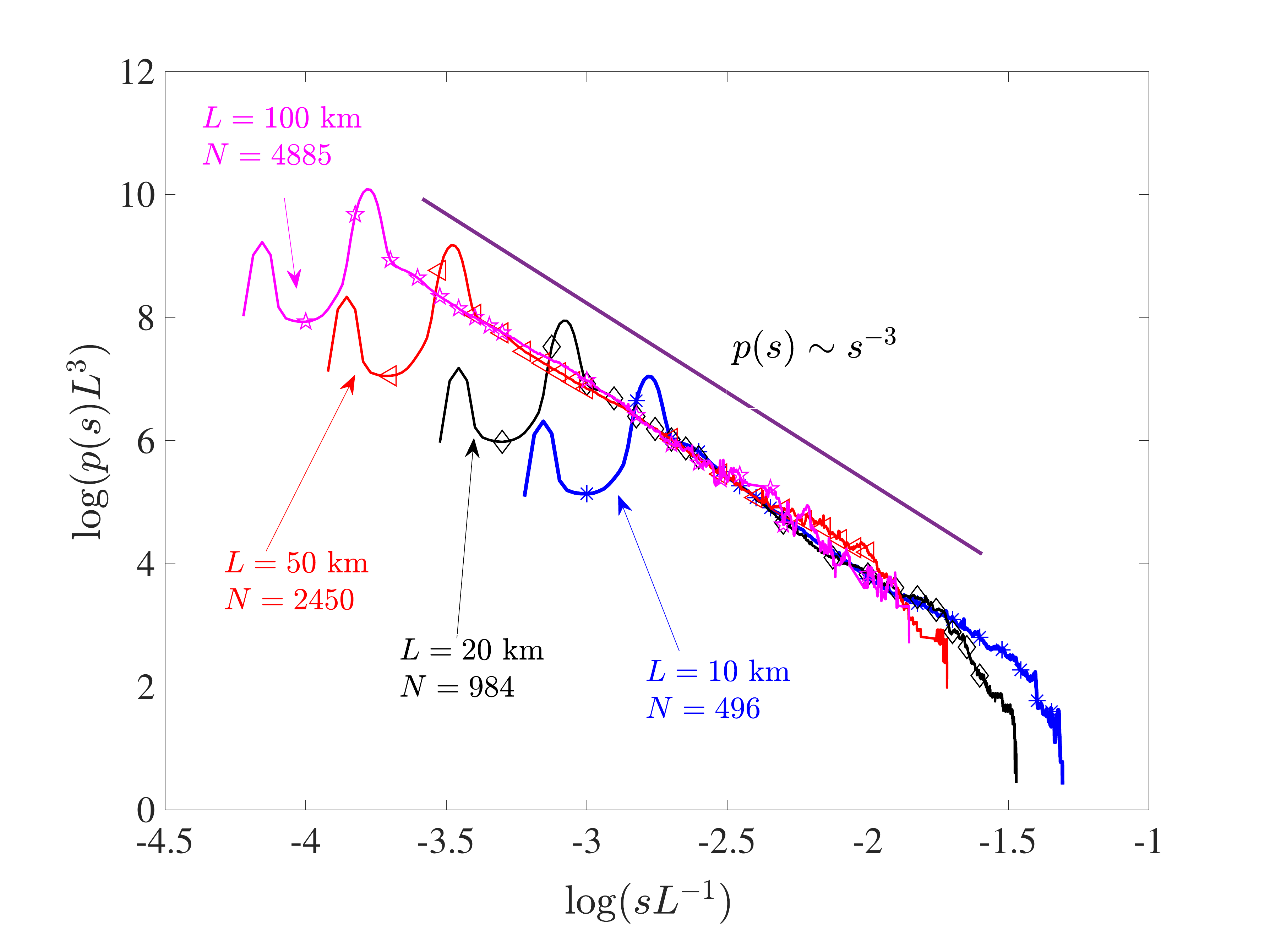}}
	\caption{Scaled gap distributions near $\rho_{\cc}$ for various $\{L,N\}$ pairs.}
	\label{fig:psvss-rt}
\end{figure}

 In addition, we observed that the gap distribution $p(s)$ decays as a power-law just above  $\rho_{\cc}$ .
 A finite size scaling plot of $p(s)$ for various $L$s is shown in Fig.~\ref{fig:psvss-rt}, from which we deduced that 
$p(s) \sim s^{-\alpha}$ with $\alpha=3$ asymptotically before the finite size effects takeover. The power-law distribution of gaps just above $\rho_{\cc}$ indicates presence of multiple length-scales in the system, which correspond to the gaps of various sizes that form between the platoons of various sizes with the largest gap being proportional to the size of the system. This is related to the power-law divergence seen in $\Delta^2$ and $\langle s^{\max} \rangle$.
 
The existence of both the power-law divergence and the discontinuity in $\langle s^{\max} \rangle$ and $\Delta^2$ in the neighborhood of the phase-transition is not reported in the traffic flow theory literature to our knowledge. The power-law divergence in $\Delta^2$ and the power-law tail of $p(s)$ have been observed in TASEP and NS models. Krug and Ferrari \cite{Krug1996} found, using heuristic arguments, that $\eta=(1-n)/n$ for $n \in (0,1)$ and it logarithmically diverges when $n=1$ where $n= a_{v_{\ff}}-1$. They also found that $\alpha= n+2$. These results have also been observed in simulations of the NS model. For the present study, $a_{v_{\ff}}=2$ and therefore, according to Krug et al., $\Delta^2$ should have diverged logarithmically but we have shown above that $\eta =1$ which is a different value. However, the value of the exponent $\alpha$ we got agrees with those predicted for TASEP and NS models. The above results and analysis show the complex nature of the transition. The power-law divergence of $\Delta^2$ as $\rho \rightarrow \rho_{\cc}^+$ resembles a second order transition. However, the infinite discontinuity in the $\Delta^2$ is unexpected of a second order transition. On the contrary, the transition may not be called first order owing to the power-law divergence of $\Delta^2$.  Thus we see the transition to be of unusual sort and its properties do not match with that observed in the NS model except for the power-law gap distribution just above $\rho_{\cc}$.


\section{Summary and Discussion}
In this work we have studied the effect of reaction time on the emergent phenomena in a heterogeneous traffic using numerical simulations of a version of Newell's model.  The heterogeneity is incorporated using quenched disorders in 
each of the parameters of the model and in the reaction time. The dynamical exponents describing the platoon forming phenomenon at low density are noted to be the same as those derived by Ben Naim et al. for ballistic aggregation and those observed in the Nagel-Schkrekenberg (NS) model. 

In the single giant platoon that forms in the stationary state at low densities, we observed that spontaneous stop-go waves form somewhere inside the platoon and move upstream until the tail of the jam is reached where they dissipate. The phase transition happens when the head of the giant platoon interacts with its tail and the stop-go waves circulate continuously all through the ring without dissipation. The transition density closely matches with the reciprocal of the expectation value of the gap distribution in the giant platoon in the thermodynamic limit and it is numerically observed to be lower than the transition density when there is no reaction time.  To understand and characterize the transition, we determined the gap distribution ($p(s)$), the maximum gap in the system ($\langle s^{\max} \rangle$) and the variance  ($\Delta^2$) of the gap distribution for various densities across the transition density $\rho_{\cc}$. 

The following picture emerges from the observations we made: First, it may be noted that the phase-transition in the present system happens because of a competition between the phase-ordering due to the quenched disorder in the free-flow speed and the formation of stable stop-go waves which destroy the phase-ordering. Below $\rho_{\cc}$, the phase ordering effect wins and the system segregates into a single giant platoon and a large gap ahead of it, $\left\langle s^{\max}\right\rangle$ simply represents the gap ahead of the platoon leader, which diverges in the thermodynamic limit. As density is increased, keeping the system size constant, the gap ahead of the leader reduces and in the limit $\rho\rightarrow\rho_{\cc}^-$ the gap ahead of the leader becomes less than their critical gap, thus $\left\langle s^{\max}\right\rangle$ becomes finite and size-independent.  This may also be seen from the fact that $p_{\gx}(s)$ merges with $p_{\pp}(s)$ as $\rho \rightarrow \rho_{\cc}^-$, thus making $p(s)$ normalizable with a finite variance $\Delta^2$. On the other hand, as the critical density is crossed ($\rho \rightarrow \rho_{\cc}^+$), the stop-go waves become dominant and obstruct the formation of a giant platoon. However, between two stop-go waves, the vehicles form clusters of various possible sizes because of the phase-ordering effect.  As the system size becomes very large, it is possible to have a large enough distance between two stop-go waves that allows the formation of large platoons and, thus, gaps proportional to the size of the system. Evidence of this happening in the system is the power-law distribution of the gaps $p(s) \sim s^{-3}$ which renders the variance $\Delta^2$ and $\langle s^{\max} \rangle$ to diverge. Thus, we have a complex situation in which the quantities $\Delta^2$ and $\langle s^{\max} \rangle$ become finite as $ \rho\rightarrow\rho_{\cc}^-$ and diverge as $\rho\rightarrow\rho_{\cc}^+$ thereby creating a discontinuity at $\rho_{\cc}$. Thus the phase transition observed here is of unusual sort with properties of both first and second order transitions. We note that such transitions with properties of both first and second order transition are observed in other systems like granular media and various other systems like foams and colloids. However, a unified picture of all these transitions is still an open question and with our present study, perhaps heterogeneous traffic joins this class of systems.\footnote{We thank an anonymous reviewer for bringing this to our attention.} 

Overall, we find that the present work reveals some novel aspects of phase transitions in heterogeneous traffic flow in the context of car-following models. Insights from our study may be useful in developing continuum theories for 
heterogeneous traffic flow, which have applications in transportation engineering and traffic management. Further, the unusual nature of the phase transition may have implications on fuel economy and pollution as there would be frequent breaking and acceleration maneuvers. Modeling of these aspects and applications that aim to avoid stop-go maneuvers are gaining traction in the engineering literature; see, e.g. \cite{ard2020microsimulation} and references therein. The power-laws concerning kinetics would give an idea about timescale of build-up of traffic on a highway.  This is of particular importance in the context of network control techniques that aim to ``stabilize'' traffic networks, which are gaining a lot of popularity in the engineering literature (see, e.g., the work of the second author on the subject \cite{Li2019position,Li2020queue}), and are even being tested in the real-world for feasibility (see, e.g., \cite{Mercader2020max}).

\begin{acknowledgments}
This work was supported by the NYUAD Center for Interacting Urban Networks (CITIES), funded by Tamkeen under the NYUAD Research Institute Award CG001 and by the Swiss Re Institute under the Quantum Cities\textsuperscript{TM} initiative.  The views expressed in this paper are those of the authors and do not reflect the opinions of CITIES or the funding agencies.
\end{acknowledgments}

\appendix
\section{Equilibration of a follower's speed to that of a slower leader}
Below we discuss, in mathematical terms, the effect of the delay induced by the reaction time ($\tau$) in the Newell's car-following model when a follower vehicle moving at a high speed equilibrates its speed with that of a slow moving leader. Denote  $v_{\ff,\phi}$, $S_{\jj,\phi}$ and $w_{\bb,\phi}$ and  $v_{\ff,\lambda}$, $S_{\jj,\lambda}$ and $w_{\bb,\lambda}$  as the free-flow speed, jam gap and the backward wave speed of the follower ($\phi$) and the leader ($\lambda$), respectively. Suppose $v_{\ff,\phi} > v_{\ff,\lambda}$ and assume that both vehicles were far apart and moving at their respective free flow (or desired) speeds, and that the follower reaches their critical gap $S_{\cc,\phi}$ at time $t_0$. The follower then begins to adapt to the speed of the leader. For simplicity, let's assume that the follower remains in the congestion regime i.e., at a gap $S_{\jj,\phi}<s<S_{\cc,\phi}$ for all $t > t_0$. The leader continues to coast at their initial speed even after $t_0$ as there is no vehicle ahead of it. Thus,
\begin{equation}
	x_{\lambda}(t) = x_{\lambda}(t_0) + (t-t_0) v_{\ff,\lambda}. \label{leom}
\end{equation}
The equation of motion of the follower is
\begin{equation}
	x_{\phi}(t) = x_{\phi}(t_0) + \int_{t_0}^t \dd t^{\prime} V\big(s_{\phi}(t^{\prime}-\tau)\big). \label{feom}
\end{equation}
Equation~\eqref{feom} can be integrated analytically in a piece-wise manner in intervals of reaction time. In the interval $[t_0,t_0+\tau)$, since $s_{\phi}(t_0-\tau)> S_{\cc,\phi}$, the follower doesn't change their speed because of the delay due to reaction time. Therefore,
\begin{equation}
	x_{\phi}(t) = x_{\phi}(t_0) + (t-t_0)v_{\ff,\phi} \label{xf1}
\end{equation}
for $t \in [t_0,t_0 + \tau)$.  Thus,
\begin{equation}
	s_{\phi}(t) = x_{\lambda}(t) - x_{\phi}(t) = S_{\cc,\phi} - (t-t_0) \delta v_{\ff} \label{sf1}
\end{equation}
for $t \in [t_0,t_0 + \tau)$, where $\delta v_{\ff} = v_{\ff,\phi} - v_{\ff,\lambda}$. In the next time interval $[t_0+\tau,t_0 + 2\tau)$, the follower starts responding to the reduction in gap in the previous time interval and reduces their speed in accord with Eq.~\eqref{vs}. Combining Eq.~\eqref{xf1} and Eq.~\eqref{sf1} with Eq.~\eqref{feom} and integrating, we get for $t \in [t_0+\tau,t_0 + 2\tau)$
\begin{equation}
		s_{\phi}(t) = S_{\cc,\phi} - (t - t_0) \delta v_{\ff} + \frac{\delta v_{\ff}}{2!} \frac{w_{\bb,\phi}}{S_{\jj,\phi}} \big(t - (t_0 + \tau)\big)^2.
	\label{sf2}
\end{equation}
Similarly, for $t \in [t_0+2\tau,t_0 + 3\tau)$:
\begin{multline}
		s_{\phi}(t) = S_{\cc,\phi} - (t - t_0) \delta v_{\ff} + \frac{\delta v_{\ff}}{2!} \frac{w_{\bb,\phi}}{S_{\jj,\phi}} \big(t - (t_o + \tau)\big)^2 \\
		- \frac{\delta v_{\ff}}{3!} \Big(\frac{w_{\bb,\phi}}{S_{\jj,\phi}}\Big)^2 \big(t - (t_0 + 2\tau)\big)^3.
	\label{sf3}
\end{multline}
In general, we obtain
\begin{multline}
	s_{\phi}(t) = S_{\cc,\phi} + \sum_{n=0}^{\infty} \bigg( \frac{(-1)^{n+1} \delta v_{\ff}}{(n+1)!} \Big(\frac{w_{\bb,\phi}}{S_{\jj,\phi}}\Big)^n \\  
	\times \big(t - (t_0 + n\tau)\big)^{n+1} I_{[t_0 + n \tau, \infty )}(t) \bigg).
	\label{sfn}
\end{multline}
Eq.~\eqref{sfn} is  essentially a polynomial with new terms added as time evolves. It can be easily checked using the ratio test that the series converges. Below we analyze the equation in the limit of small $\tau$ and obtain some insights regarding the effect of reaction time.

First, we investigate the limiting (in time) behavior of $s_{\phi}$ when $\tau=0$. When $\tau = 0$, we have that
\begin{equation}
		s_{\phi}(t) = S_{\cc,\phi} + (t - t_0) v_{\ff,\lambda} - \int_{t_0}^t \dd t^{\prime} V\big( s_{\phi}(t^{\prime}) \big).
	\label{sfntau0_1}
\end{equation}
Since $s_{\phi}(t) < S_{\cc,\phi}$ for $t > t_0$, it can be shown that Eq.~\eqref{sfntau0_1} has the following solution: 
\begin{multline}
		s_{\phi}(t) = S_{\cc,\phi} \mathrm{e}^{-\frac{w_{\bb,\phi}}{S_{\jj,\phi}}(t-t_0)} \\
		+ \frac{S_{\jj,\phi}}{w_{\bb,\phi}} (v_{\ff,\lambda} + w_{\bb,\phi}) \Big( 1 - \mathrm{e}^{-\frac{w_{\bb,\phi}}{S_{\jj,\phi}}(t-t_0)} \Big),
	\label{sfntau0_2}
\end{multline}
which tends to the equilibrium gap exactly as dictated by Eq.~\eqref{vs} in the long time limit. This is also the case in Eq.~\eqref{sfn}, when $\tau \rightarrow 0$. To demonstrate this, we first assume without loss of generality that $t_0 = 0$ and write Eq.~\eqref{sfn} as
\begin{equation}
		s_{\phi}(t) = S_{\cc,\phi} + B \sum_{n=1}^{\infty} \frac{(-1)^n}{n!} A^n \big(t - (n-1) \tau\big)^n, \label{sfn_1}
\end{equation}
where $A \equiv w_{\bb,\phi}S_{\jj,\phi}^{-1}$ and $B \equiv \delta v_{\ff} A^{-1}$.  As $\tau \rightarrow 0$
\begin{equation}
	s_{\phi}(t) \rightarrow S_{\cc,\phi} + \big( \mathrm{e}^{- A t} - 1\big) B, \label{sfntau2}
\end{equation}
which converges to the equilibrium gap given by Eq.~\eqref{vs} in the long time limit. Eq.~\eqref{sfntau0_2} indicates the presence of a characteristic time scale $S_{\jj,\phi} w_{\bb,\phi}^{-1}$ for relaxation of the speed of the 
follower to that of the leader. This should  imply that if $\tau \ll S_{\jj,\phi} w_{\bb,\phi}^{-1}$, the reaction time will have little to no effect on the speed equilibration process.  To see this, we approximate the series Eq.~\eqref{sfn_1} as follows
\begin{multline}
	s_{\phi}(t) = S_{\cc,\phi} + B \sum_{n=1}^{\infty} \frac{(-A)^n}{n!} \sum_{r=0}^n \binom{n}{r} t^{n-r} (-(n-1)\tau)^r \\
	\approx S_{\cc,\phi} + B \sum_{n=1}^{\infty} \frac{(-A)^n}{n!} \Big( t^n - n(n-1)t^{n-1} \tau \\
	+ \frac{n(n-1)}{2} t^{n-2} (n-1)^2 \tau^2 \Big),
	\label{sfnpt}
\end{multline}
where we have truncated the inner binomial expansion after the second term. Next, using the series expansion of exponential and after some algebra, we get
\begin{multline}
		s_{\phi}(t) \approx S_{\cc,\phi} - B \\
	+ B \Big( 1 - tA^2 \tau + \frac{1}{2} A^2 \tau^2 \big( A^2t^2 - 3tA \big) \Big) \mathrm{e}^{-At}
\label{sfntau3}
\end{multline}
\begin{figure}[h!]
	\centering
	\resizebox{0.45\textwidth}{!}{%
		\includegraphics[]{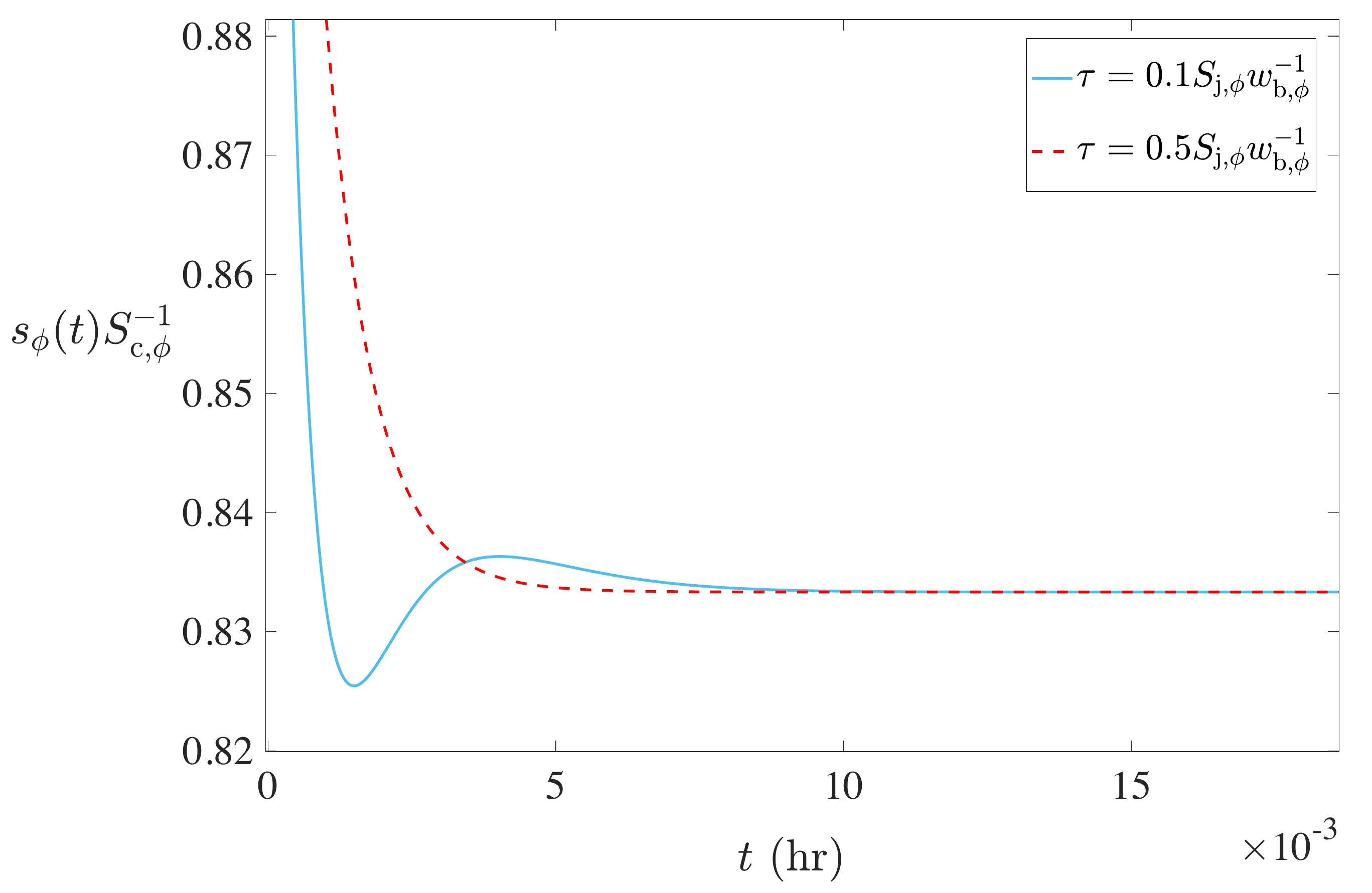}}
	\caption{Gap ahead of the follower ($\phi$) as a function of time for a typical case calculated using Eq.~\eqref{sfntau3}. For small $\tau$, the relaxation to stationary gap is monotonic. As $\tau$ approaches $S_{\jj,\phi} w_{\bb,\phi}^{-1}$,	$s_{\phi}(t)$ becomes non-monotone.}
	\label{fig:S_pt}
\end{figure}
The above expression for $s_{\phi}$ clearly illustrates the effect of a non-zero reaction time; see also Fig.~\ref{fig:S_pt}. For very small reaction time $\tau \ll S_{\jj} w_{\bb,\phi}^{-1}$, the quadratic polynomial multiplying the exponential term in Eq.~\eqref{sfntau3} has no real roots and converges monotonically to the stationary value $S_{\cc,\phi} - B$.  In fact Eq.~\eqref{sfntau3} tends to Eq.~\eqref{sfntau2} in the limit $\tau \rightarrow 0$.  As $\tau$ is increased slightly, the quadratic polynomial begins to have positive real roots and thus $s_{\phi}$ becomes non-monotonic with its value overshooting and undershooting the stationary gap $S_{\cc,\phi} - B$ at various instances before finally relaxing to this value. As $\tau \sim S_{\jj,\phi} w_{\bb,\phi}^{-1}$, the  higher order terms in the series become dominant resulting in more vigorous oscillations which decay slowly.


\bibliographystyle{unsrtnat}
\bibliography{aipsamp}

\begin{thebibliography}{49}
\providecommand{\natexlab}[1]{#1}
\providecommand{\url}[1]{\texttt{#1}}
\expandafter\ifx\csname urlstyle\endcsname\relax
  \providecommand{\doi}[1]{doi: #1}\else
  \providecommand{\doi}{doi: \begingroup \urlstyle{rm}\Url}\fi

\bibitem[Helbing(2001)]{Helbing2001}
Dirk Helbing.
\newblock Traffic and related self-driven many-particle systems.
\newblock \emph{Reviews of modern physics}, 73\penalty0 (4):\penalty0 1067,
  2001.

\bibitem[Chowdhury et~al.(2000{\natexlab{a}})Chowdhury, Santen, and
  Schadschneider]{DC2000}
Debashish Chowdhury, Ludger Santen, and Andreas Schadschneider.
\newblock Statistical physics of vehicular traffic and some related systems.
\newblock \emph{Physics Reports}, 329\penalty0 (4-6):\penalty0 199--329,
  2000{\natexlab{a}}.

\bibitem[Nagel and Schreckenberg(1992)]{stocNS1992}
Kai Nagel and Michael Schreckenberg.
\newblock A cellular automaton model for freeway traffic.
\newblock \emph{Journal de physique I}, 2\penalty0 (12):\penalty0 2221, 1992.

\bibitem[Nagel and Herrmann(1993)]{Nagel1993}
Kai Nagel and Hans~J Herrmann.
\newblock Deterministic models for traffic jams.
\newblock \emph{Physica A: Statistical Mechanics and its Applications},
  199\penalty0 (2):\penalty0 254--269, 1993.

\bibitem[Chowdhury et~al.(1997)Chowdhury, Ghosh, Majumdar, Sinha, and
  Stinchcombe]{Chowdhury1997}
Debashish Chowdhury, Kingshuk Ghosh, Arnab Majumdar, Shishir Sinha, and
  RB~Stinchcombe.
\newblock Particle-hopping models of vehicular traffic: Distributions of
  distance headways and distance between jams.
\newblock \emph{Physica A: Statistical Mechanics and its Applications},
  246\penalty0 (3-4):\penalty0 471--486, 1997.

\bibitem[Nagel et~al.(2000)Nagel, Esser, and Rickert]{Nagel}
Kai Nagel, J\"org Esser, and Marcus Rickert.
\newblock \emph{Large-scale traffic simulations for transportation planning},
  pages 151--202.
\newblock World Scientific, Singapore, 2000.
\newblock \doi{10.1142/9789812813329_0006}.

\bibitem[Behrisch et~al.(2011)Behrisch, Bieker, Erdmann, and Krajzewicz]{Sumo}
Michael Behrisch, Laura Bieker, Jakob Erdmann, and Daniel Krajzewicz.
\newblock Sumo--simulation of urban mobility: an overview.
\newblock In \emph{Proceedings of SIMUL 2011, The Third International
  Conference on Advances in System Simulation}. ThinkMind, 2011.

\bibitem[Fellendorf and Vortisch(2010)]{Vissim}
Martin Fellendorf and Peter Vortisch.
\newblock Microscopic traffic flow simulator vissim.
\newblock In \emph{Fundamentals of traffic simulation}, pages 63--93. Springer,
  2010.

\bibitem[Ramana and Jabari(2020)]{Ramana2020}
A~Sai~Venkata Ramana and Saif~Eddin Jabari.
\newblock Traffic flow with multiple quenched disorders.
\newblock \emph{Physical Review E}, 101\penalty0 (5):\penalty0 052127, 2020.

\bibitem[Cs{\'a}nyi and Kert{\'e}sz(1995)]{Csanyi1995}
G~Cs{\'a}nyi and J~Kert{\'e}sz.
\newblock Scaling behaviour in discrete traffic models.
\newblock \emph{Journal of Physics A: Mathematical and General}, 28\penalty0
  (16):\penalty0 L427, 1995.

\bibitem[Sasv{\'a}ri and Kert{\'e}sz(1997)]{Sasvari1997}
M{\'a}rton Sasv{\'a}ri and J{\'a}nos Kert{\'e}sz.
\newblock Cellular automata models of single-lane traffic.
\newblock \emph{Physical Review E}, 56\penalty0 (4):\penalty0 4104, 1997.

\bibitem[de~Gier et~al.(2019)de~Gier, Schadschneider, Schmidt, and
  Sch{\"u}tz]{Gier2019}
Jan de~Gier, Andreas Schadschneider, Johannes Schmidt, and Gunter~M Sch{\"u}tz.
\newblock Kardar-parisi-zhang universality of the nagel-schreckenberg model.
\newblock \emph{Physical Review E}, 100\penalty0 (5):\penalty0 052111, 2019.

\bibitem[L{\"u}beck et~al.(1998)L{\"u}beck, Schreckenberg, and
  Usadel]{Lubeck1998}
S~L{\"u}beck, M~Schreckenberg, and KD~Usadel.
\newblock Density fluctuations and phase transition in the nagel-schreckenberg
  traffic flow model.
\newblock \emph{Physical Review E}, 57\penalty0 (1):\penalty0 1171, 1998.

\bibitem[Roters et~al.(1999)Roters, L{\"u}beck, and Usadel]{Roters1999}
L~Roters, S~L{\"u}beck, and KD~Usadel.
\newblock Critical behavior of a traffic flow model.
\newblock \emph{Physical Review E}, 59\penalty0 (3):\penalty0 2672, 1999.

\bibitem[Chowdhury et~al.(2000{\natexlab{b}})Chowdhury, Kert{\'e}sz, Nagel,
  Santen, and Schadschneider]{DC2000comment}
D~Chowdhury, J~Kert{\'e}sz, K~Nagel, L~Santen, and A~Schadschneider.
\newblock Comment on “critical behavior of a traffic flow model”.
\newblock \emph{Physical Review E}, 61\penalty0 (3):\penalty0 3270,
  2000{\natexlab{b}}.

\bibitem[Roters et~al.(2000)Roters, L{\"u}beck, and Usadel]{Roters2000reply}
L~Roters, S~L{\"u}beck, and KD~Usadel.
\newblock Reply to “comment on ‘critical behavior of a traffic flow
  model’”.
\newblock \emph{Physical Review E}, 61\penalty0 (3):\penalty0 3272, 2000.

\bibitem[Eisenbl{\"a}tter et~al.(1998)Eisenbl{\"a}tter, Santen, Schadschneider,
  and Schreckenberg]{Eisenblatter1998}
B~Eisenbl{\"a}tter, L~Santen, A~Schadschneider, and M~Schreckenberg.
\newblock Jamming transition in a cellular automaton model for traffic flow.
\newblock \emph{Physical Review E}, 57\penalty0 (2):\penalty0 1309, 1998.

\bibitem[Bain et~al.(2016)Bain, Emig, Ulm, and Schreckenberg]{Bain2016}
Nicolas Bain, Thorsten Emig, Franz-Josef Ulm, and Michael Schreckenberg.
\newblock Velocity statistics of the nagel-schreckenberg model.
\newblock \emph{Physical Review E}, 93\penalty0 (2):\penalty0 022305, 2016.

\bibitem[Souza and Vilar(2009)]{Souza2009}
Andr{\'e} Maur{\'\i}cio Concei{\c{c}}{\~a}o~de Souza and LCQ Vilar.
\newblock Traffic-flow cellular automaton: Order parameter and its conjugated
  field.
\newblock \emph{Physical Review E}, 80\penalty0 (2):\penalty0 021105, 2009.

\bibitem[Gerwinski and Krug(1999)]{Gerwinski1999}
M~Gerwinski and J~Krug.
\newblock Analytic approach to the critical density in cellular automata for
  traffic flow.
\newblock \emph{Physical Review E}, 60\penalty0 (1):\penalty0 188, 1999.

\bibitem[Bette et~al.(2017)Bette, Habel, Emig, and Schreckenberg]{Bette2017}
Henrik~M Bette, Lars Habel, Thorsten Emig, and Michael Schreckenberg.
\newblock Mechanisms of jamming in the nagel-schreckenberg model for traffic
  flow.
\newblock \emph{Physical Review E}, 95\penalty0 (1):\penalty0 012311, 2017.

\bibitem[Krug and Ferrari(1996)]{Krug1996}
Joachim Krug and Pablo~A Ferrari.
\newblock Phase transitions in driven diffusive systems with random rates.
\newblock \emph{Journal of Physics A: Mathematical and General}, 29\penalty0
  (18):\penalty0 L465, 1996.

\bibitem[Krug(2000)]{Krug2000}
Joachim Krug.
\newblock Phase separation in disordered exclusion models.
\newblock \emph{Brazilian Journal of Physics}, 30\penalty0 (1):\penalty0
  97--104, 2000.

\bibitem[Evans(1996)]{Evans1996}
MR~Evans.
\newblock Bose-einstein condensation in disordered exclusion models and
  relation to traffic flow.
\newblock \emph{EPL (Europhysics Letters)}, 36\penalty0 (1):\penalty0 13, 1996.

\bibitem[Ktitarev et~al.(1997)Ktitarev, Chowdhury, and Wolf]{Ktitarev1997}
Dmitri~V Ktitarev, Debashish Chowdhury, and Dietrich~E Wolf.
\newblock Stochastic traffic model with random deceleration probabilities:
  queueing and power-law gap distribution.
\newblock \emph{Journal of Physics A: Mathematical and General}, 30\penalty0
  (8):\penalty0 L221, 1997.

\bibitem[Bengrine et~al.(1999)Bengrine, Benyoussef, Ez-Zahraouy, Krug, Loulidi,
  and Mhirech]{Bengrine1999}
M~Bengrine, A~Benyoussef, H~Ez-Zahraouy, J~Krug, M~Loulidi, and F~Mhirech.
\newblock A simulation study of an asymmetric exclusion model with open
  boundaries and random rates.
\newblock \emph{Journal of Physics A: Mathematical and General}, 32\penalty0
  (13):\penalty0 2527, 1999.

\bibitem[Newell(2002)]{Newell2002}
Gordon~Frank Newell.
\newblock A simplified car-following theory: a lower order model.
\newblock \emph{Transportation Research Part B: Methodological}, 36\penalty0
  (3):\penalty0 195--205, 2002.

\bibitem[Ahn et~al.(2004)Ahn, Cassidy, and Laval]{ahn2004verification}
Soyoung Ahn, Michael~J Cassidy, and Jorge Laval.
\newblock Verification of a simplified car-following theory.
\newblock \emph{Transportation Research Part B: Methodological}, 38\penalty0
  (5):\penalty0 431--440, 2004.

\bibitem[Chiabaut et~al.(2009)Chiabaut, Buisson, and
  Leclercq]{chiabaut2009fundamental}
Nicolas Chiabaut, Christine Buisson, and Ludovic Leclercq.
\newblock Fundamental diagram estimation through passing rate measurements in
  congestion.
\newblock \emph{IEEE Transactions on Intelligent Transportation Systems},
  10\penalty0 (2):\penalty0 355--359, 2009.

\bibitem[Chiabaut et~al.(2010)Chiabaut, Leclercq, and
  Buisson]{chiabaut2010heterogeneous}
Nicolas Chiabaut, Ludovic Leclercq, and Christine Buisson.
\newblock From heterogeneous drivers to macroscopic patterns in congestion.
\newblock \emph{Transportation Research Part B: Methodological}, 44\penalty0
  (2):\penalty0 299--308, 2010.

\bibitem[Jabari et~al.(2014)Jabari, Zheng, and Liu]{jabari2014}
Saif~Eddin Jabari, Jianfeng Zheng, and Henry~X Liu.
\newblock A probabilistic stationary speed--density relation based on
  newell’s simplified car-following model.
\newblock \emph{Transportation Research Part B: Methodological}, 68:\penalty0
  205--223, 2014.

\bibitem[Jabari et~al.(2018)Jabari, Zheng, Liu, and
  Filipovska]{jabari2018stochastic}
Saif~Eddin Jabari, Fangfang Zheng, Henry~X Liu, and Monika Filipovska.
\newblock Stochastic {L}agrangian modeling of traffic dynamics.
\newblock In \emph{The 97th Annual Meeting of the Transportation Research
  Board, Washington D.C}, pages 18--04170, 2018.

\bibitem[Zheng et~al.(2018)Zheng, Jabari, Liu, and Lin]{zheng2018stochastic}
Fangfang Zheng, Saif~Eddin Jabari, Henry~X. Liu, and DianChao Lin.
\newblock Traffic state estimation using stochastic {L}agrangian dynamics.
\newblock \emph{Transportation Research Part B: Methodological}, 115:\penalty0
  143--165, 2018.

\bibitem[Jabari et~al.(2020)Jabari, Freris, and Dilip]{jabari2020sparse}
Saif~Eddin Jabari, Nikolaos~M Freris, and Deepthi~Mary Dilip.
\newblock Sparse travel time estimation from streaming data.
\newblock \emph{Transportation Science}, 54\penalty0 (1):\penalty0 1--20, 2020.

\bibitem[Treiber and Kesting(2013)]{Treiber2013}
Martin Treiber and Arne Kesting.
\newblock Traffic flow dynamics.
\newblock \emph{Traffic Flow Dynamics: Data, Models and Simulation,
  Springer-Verlag Berlin Heidelberg}, 2013.

\bibitem[Kerner(2004)]{Kerner1}
Boris~S Kerner.
\newblock \emph{The physics of traffic: {E}mpirical freeway pattern features,
  engineering applications, and theory}.
\newblock Springer-Verlag Berlin Heidelberg, 2004.

\bibitem[Kerner(2017)]{Kerner2}
Boris~S Kerner.
\newblock \emph{Breakdown in Traffic Networks}.
\newblock Springer-Verlag Berlin Heidelberg, 2017.

\bibitem[Kerner(2021)]{kerner2021effect}
Boris~S Kerner.
\newblock Effect of autonomous driving on traffic breakdown in mixed traffic
  flow: A comparison of classical acc with three-traffic-phase-acc (tpacc).
\newblock \emph{Physica A: Statistical Mechanics and its Applications},
  562:\penalty0 125315, 2021.

\bibitem[Sch{\"o}nhof and Helbing(2007)]{Schonhof2007tpt}
Martin Sch{\"o}nhof and Dirk Helbing.
\newblock Empirical features of congested traffic states and their implications
  for traffic modeling.
\newblock \emph{Transportation Science}, 41\penalty0 (2):\penalty0 135--166,
  2007.

\bibitem[Sch{\"o}nhof and Helbing(2009)]{Schonhof2009tpt}
Martin Sch{\"o}nhof and Dirk Helbing.
\newblock Criticism of three-phase traffic theory.
\newblock \emph{Transportation Research Part B: Methodological}, 43\penalty0
  (7):\penalty0 784--797, 2009.

\bibitem[Treiber et~al.(2010)Treiber, Kesting, and Helbing]{Treiber2010tpt}
Martin Treiber, Arne Kesting, and Dirk Helbing.
\newblock Three-phase traffic theory and two-phase models with a fundamental
  diagram in the light of empirical stylized facts.
\newblock \emph{Transportation Research Part B: Methodological}, 44\penalty0
  (8-9):\penalty0 983--1000, 2010.

\bibitem[Ben-Naim et~al.(1994)Ben-Naim, Krapivsky, and Redner]{Ben1994}
Eli Ben-Naim, Pavel~L Krapivsky, and Sidney Redner.
\newblock Kinetics of clustering in traffic flows.
\newblock \emph{Physical Review E}, 50\penalty0 (2):\penalty0 822, 1994.

\bibitem[Tripathy and Barma(1997)]{Tripathy1997}
Goutam Tripathy and Mustansir Barma.
\newblock Steady state and dynamics of driven diffusive systems with quenched
  disorder.
\newblock \emph{Physical review letters}, 78\penalty0 (16):\penalty0 3039,
  1997.

\bibitem[Barma(2006)]{Barma2006}
Mustansir Barma.
\newblock Driven diffusive systems with disorder.
\newblock \emph{Physica A: Statistical Mechanics and its Applications},
  372\penalty0 (1):\penalty0 22--33, 2006.

\bibitem[Balouchi and Browne(2016)]{Balouchi2016}
Ashkan Balouchi and Dana~A Browne.
\newblock Finite-size effects in the nagel-schreckenberg traffic model.
\newblock \emph{Physical Review E}, 93\penalty0 (5):\penalty0 052302, 2016.

\bibitem[Ard et~al.(2020)Ard, Dollar, Vahidi, Zhang, and
  Karbowski]{ard2020microsimulation}
Tyler Ard, Robert~Austin Dollar, Ardalan Vahidi, Yaozhong Zhang, and Dominik
  Karbowski.
\newblock Microsimulation of energy and flow effects from optimal automated
  driving in mixed traffic.
\newblock \emph{Transportation Research Part C: Emerging Technologies},
  120:\penalty0 102806, 2020.

\bibitem[Li and Jabari(2019)]{Li2019position}
Li~Li and Saif~Eddin Jabari.
\newblock Position weighted backpressure intersection control for urban
  networks.
\newblock \emph{Transportation Research Part B: Methodological}, 128:\penalty0
  435--461, 2019.

\bibitem[Li et~al.(2021)Li, Okoth, and Jabari]{Li2020queue}
Li~Li, Victor Okoth, and Saif~Eddin Jabari.
\newblock Backpressure control with estimated queue lengths for urban network
  traffic.
\newblock \emph{IET Intelligent Transport Systems}, 15\penalty0 (2):\penalty0
  320--330, 2021.

\bibitem[Mercader et~al.(2020)Mercader, Uwayid, and Haddad]{Mercader2020max}
Pedro Mercader, Wasim Uwayid, and Jack Haddad.
\newblock Max-pressure traffic controller based on travel times: An
  experimental analysis.
\newblock \emph{Transportation Research Part C: Emerging Technologies},
  110:\penalty0 275--290, 2020.

\end{thebibliography}

\end{document}